\documentclass[11pt]{article}
\usepackage{epsf}
\usepackage{axodraw}

\newcommand{ \centeron }[2]{{\setbox0=\hbox{#1}\setbox1=\hbox{#2}\ifdim
                             \wd1>\wd0\kern.5\wd1\kern-.5\wd0\fi \copy0
                             \kern-.5\wd0\kern-.5\wd1\copy1\ifdim\wd0>\wd1
                             \kern.5\wd0\kern-.5\wd1\fi}}
\newcommand{ \ltap }{\>\centeron{\raise.35ex\hbox{$<$}}
                     {\lower.65ex\hbox{$\sim$}}\>}
\newcommand{ \gtap }{\>\centeron{\raise.35ex\hbox{$>$}}
                     {\lower.65ex\hbox{$\sim$}}\>}
\newcommand{ \gsim }{\mathrel{\gtap}}
\newcommand{ \lsim }{\mathrel{\ltap}}

\newcommand{ \slashchar }[1]{\setbox0=\hbox{$#1$}   
   \dimen0=\wd0                                     
   \setbox1=\hbox{/} \dimen1=\wd1                   
   \ifdim\dimen0>\dimen1                            
      \rlap{\hbox to \dimen0{\hfil/\hfil}}          
      #1                                            
   \else                                            
      \rlap{\hbox to \dimen1{\hfil$#1$\hfil}}       
      /                                             
   \fi}                                             %

\newcommand{ \ra       }{\rightarrow}
\newcommand{ \textfrac }[2]{ {\textstyle\frac{#1}{#2}} }
\newcommand{ \ph       }{\gamma}

\newcommand{ \snu       }{\tilde{\nu}}
\newcommand{ \snue      }{\tilde{\nu}_e}
\newcommand{ \snumu     }{\tilde{\nu}_\mu}

\newcommand{ \snuone    }{\tilde{\nu}_1}
\newcommand{ \snutwo    }{\tilde{\nu}_2}

\newcommand{ \snuonetwo }{\tilde{\nu}_{1,2}}

\newcommand{ \se        }{\tilde{e}}
\newcommand{ \smu       }{\tilde{\mu}}

\newcommand{ \sll       }{\tilde{\ell}} 

\newcommand{ \cv        }{\cos\theta_{\tilde{\nu}}}
\newcommand{ \sv        }{\sin\theta_{\tilde{\nu}}}
\newcommand{ \stwov     }{\sin \, 2\theta_{\tilde{\nu}}}
\newcommand{ \twosv     }{\sin^2 \theta_{\tilde{\nu}}}
\newcommand{ \twocv     }{\cos^2 \theta_{\tilde{\nu}}}
\newcommand{ \thetav    }{\theta_{\tilde{\nu}}}

\newcommand{ \Cpm       }{\tilde{\chi}^\pm}
\newcommand{ \N         }{\tilde{\chi}^0}

\newcommand{ \Bino      }{\tilde{B}}
\newcommand{ \Wino      }{\tilde{W}}
\newcommand{ \Higgsino  }{\tilde{H}}

\newcommand{ \Pl        }{\mbox{Pl}}

\def\singleandabitspaced{\baselineskip=\normalbaselineskip\multiply
    \baselineskip by 110\divide\baselineskip by 100}
\def\abstractspacing{\baselineskip=\normalbaselineskip\multiply
    \baselineskip by 110\divide\baselineskip by 100}
\def\singlespaced{\baselineskip=\normalbaselineskip}

\begin{document}

\singlespaced

\begin{titlepage}

\begin{flushright}
hep-ph/0104317 \\
MADPH--01--1225 \\
UW/PT-01-12 \\
\end{flushright}

\vspace{1.5cm}

\begin{center}
\mbox{\Large \textbf{Constraints on lepton flavor violation in 
      the MSSM}} \\
\vspace*{0.3cm}
\mbox{\Large \textbf{from the muon anomalous magnetic moment measurement}}
\\

\vspace*{2.0cm}
{\Large Z. Chacko$^{(1)}$ and Graham D. Kribs$^{(2)}$} \\
\vspace*{0.5cm}
\textit{$^{(1)}$Department of Physics, Box 31560, University of Washington, \\
        Seattle, WA~~98195} \\
\vspace*{0.5cm}
\textit{$^{(2)}$Department of Physics, University of Wisconsin, \\
        1150 University Ave., Madison, WI~~53706-1390} \\

\vspace*{1.0cm}

\texttt{zchacko@phys.washington.edu, kribs@pheno.physics.wisc.edu}

\vspace*{1.0cm}

\begin{abstract}
\indent

\abstractspacing

We establish a  correspondence between those Feynman
diagrams in the MSSM which give supersymmetric contributions to
the muon anomalous magnetic moment and those which contribute to the
flavor violating processes $\mu \ra e\ph$
and $\tau \ra \mu\ph$.  Using current experimental limits on the
branching ratios of these decay modes, combined with the assumption of a
supersymmetric
contribution to the muon anomalous magnetic moment, we establish
bounds on the size of the lepton flavor violating soft masses in
the MSSM largely independent of assumptions about
other supersymmetric parameters. If the deviation measured at Brookhaven
National Laboratory 
is from supersymmetry, we find the bounds ${m^2}_{e \mu}/ {{\bar{m}}^2}
\lsim
2 \times 10^{-4}$ and
${m^2}_{\tau \mu}/ {{\bar{m}}^2} \lsim  1 \times 10^{-1}$, where
${\bar{m}}^2$ is the mass of the heaviest particle in any loop that
contributes at this level to the anomalous magnetic moment of the muon.
This provides a significant constraint
on the non-flavor-blind mediation of supersymmetry breaking that often
occurs at a suppressed level in many models, including gaugino mediation.

\end{abstract}

\end{center}
\end{titlepage}

\newpage
\setcounter{page}{2}
\renewcommand{\thefootnote}{\arabic{footnote}}
\setcounter{footnote}{0}
\singleandabitspaced

\section{Introduction}
\label{introduction-sec}

The recent measurement \cite{g-2experiment} of the anomalous 
magnetic moment (g - 2) of the muon at BNL exhibits a deviation from 
what is expected from the Standard Model (SM) (for a clear review,
see Ref.~\cite{CM}).  This suggests that 
a contribution from physics beyond the SM
is necessary to explain the discrepancy.  Supersymmetry has
been known for some time to provide a significant contribution
to anomalous magnetic moment operators \cite{old-mag-moment,Moroi,CGW,MW}.
Several recent papers have also 
considered the possibility that the BNL measurement is evidence of a
supersymmetric contribution to (g - 2) at the level of the 
experiment \cite{SUSY}.  Other papers have also have considered
a possible connection between muon (g - 2) and lepton flavor
violation in both a supersymmetric \cite{SUSYlfv} and 
nonsupersymmetric context \cite{lfv}.
In this paper we establish a correspondence between the supersymmetric
diagrams in the MSSM that contribute to the anomalous magnetic moment of
the muon and a precisely analogous class of diagrams that contribute 
to lepton flavor violating processes including 
$\mu \ra e \gamma$ and $\tau \ra \mu \gamma$. 
This enables us to place bounds on the flavor violating soft
masses ${m^2}_{e \mu}$ and ${m^2}_{\tau \mu}$ in the MSSM. In particular
we find ${m^2}_{e \mu}/ {{\bar{m}}^2} \le 2 \times 10^{-4}$ and
${m^2}_{\tau \mu}/ {{\bar{m}}^2} \le 1 \times 10^{-1}$, where
${\bar{m}}^2$ is the mass of the heaviest supersymmetric particle in any
loop that
contributes at this level to the anomalous magnetic moment of the muon.
We do not assume any relation between the gaugino soft masses  
$M_1$, $M_2$, and the $\mu$ term in the superpotential,
nor any relation among the slepton masses.  The bounds we
establish are largely insensitive to any supersymmetric parameters
and assume only that there is no accidental cancellation between 
independent diagrams contributing to the flavor violating processes.

To understand the underlying reason for the correspondence between the  
diagrams for the two types of processes consider the structure of the
relevant operators.  The anomalous magnetic moment operator has the form
\begin{eqnarray}
{\cal M}_\mu &=& \frac{i e}{2 m_\mu}
                 \overline{u}_\mu(p_2) a_\mu
                 \sigma^{\mu\nu} q_{\nu} u_\mu(p_1) A_\mu \; .
\end{eqnarray}
The operator for the process $\mu \ra e\ph$ has the form
\begin{eqnarray}
{\cal M}_{\mu e \ph} &=& \frac{i e}{2 m_\mu}
             \overline{u}_e(p_2) \sigma^{\mu\nu} q_{\nu}
             \left( a_l P_L + a_r P_R \right) u_\mu(p_1) A_\mu + h.c. \; ,
\label{mag-moment-eq}
\end{eqnarray}
where $P_{L,R} \equiv (1 \mp \gamma_5)/2$ and
$\sigma^{\mu\nu} \equiv \textfrac{i}{2} \left[ \gamma^{\mu},
\gamma^{\nu} \right]$.  
This has the same structure as the anomalous magnetic moment
operator above provided $a_L = a_R$. In particular both operators involve
a net chirality flip between the ingoing and outgoing leptons. This
suggests that the different sets of graphs contributing to the two
processes will have almost identical structures.

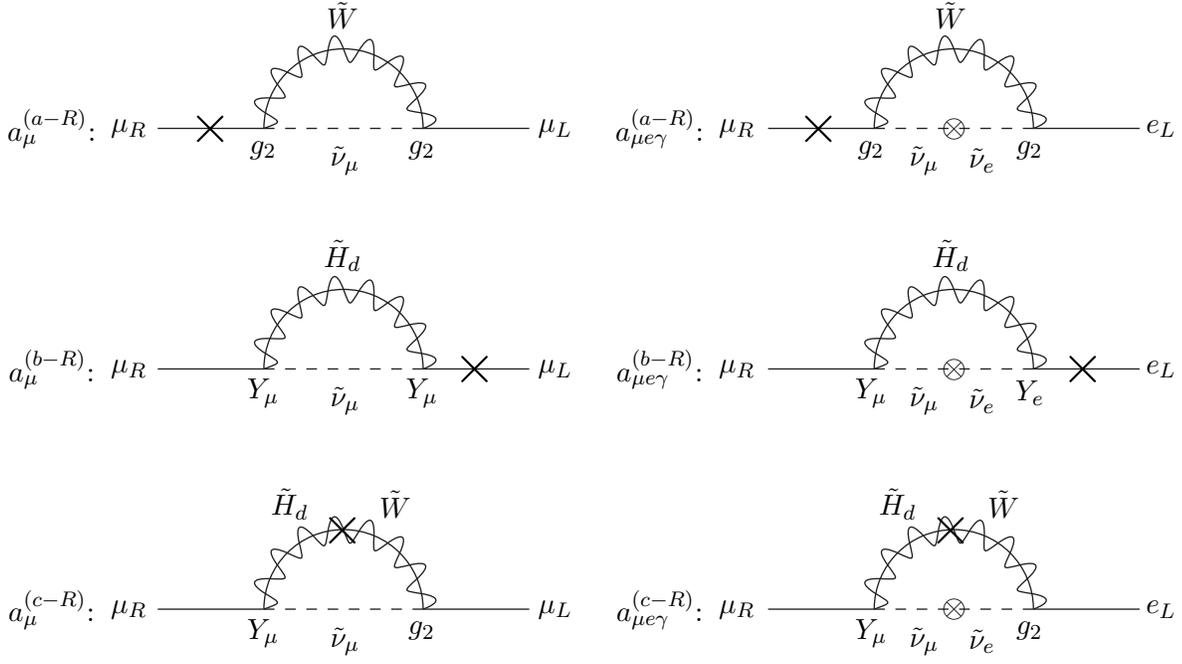
\begin{figure}[t]
\begin{picture}(440,90)
%
  \DashLine(    80, 40 )( 140, 40 ){4}
  \PhotonArc(  110, 40 )( 30, 0, 180 ){5}{8}
  \CArc(       110, 40 )( 30, 0, 180 )
  \Line(  40, 40 )( 80, 40 )
  \Line( 140, 40 )( 180, 40 )
  \Text(  60, 40 )[c]{\huge $\times$}
  \Text(  36, 40 )[r]{$\mu_R$}
  \Text( 184, 40 )[l]{$\mu_L$}
  \Text(  80, 32 )[c]{$g_2$}
  \Text( 140, 32 )[c]{$g_2$}
  \Text( 110, 83 )[c]{$\Wino$}
  \Text( 111, 28 )[c]{$\snumu$}
  \Text(   0, 40 )[c]{$a_\mu^{(a-R)}$:}
%
  \DashLine(   310, 40 )( 370, 40 ){4}
  \PhotonArc(  340, 40 )( 30, 0, 180 ){5}{8}
  \CArc(       340, 40 )( 30, 0, 180 )
  \Line( 270, 40 )( 310, 40 )
  \Line( 370, 40 )( 410, 40 )
  \Text( 290, 40 )[c]{\huge $\times$}
  \Text( 266, 40 )[r]{$\mu_R$}
  \Text( 414, 40 )[l]{$e_L$}
  \Text( 310, 32 )[c]{$g_2$}
  \Text( 370, 32 )[c]{$g_2$}
  \Text( 340, 83 )[c]{$\Wino$}
  \Text( 341.3, 40 )[c]{{\large $\otimes$}}
  \Text( 330, 28 )[c]{$\snumu$}
  \Text( 352, 28 )[c]{$\snue$}
  \Text( 230, 40 )[c]{$a_{\mu e \ph}^{(a-R)}$:}
\end{picture}
\begin{picture}(440,90)
%
  \DashLine(    80, 40 )( 140, 40 ){4}
  \PhotonArc(  110, 40 )( 30, 0, 180 ){5}{8}
  \CArc(       110, 40 )( 30, 0, 180 )
  \Line(  40, 40 )( 80, 40 )
  \Line( 140, 40 )( 180, 40 )
  \Text( 160, 40 )[c]{\huge $\times$}
  \Text(  36, 40 )[r]{$\mu_R$}
  \Text( 184, 40 )[l]{$\mu_L$}
  \Text(  80, 31 )[c]{$Y_\mu$}
  \Text( 140, 31 )[c]{$Y_\mu$}
  \Text( 110, 83 )[c]{$\Higgsino_d$}
  \Text( 111, 28 )[c]{$\snumu$}
  \Text(   0, 40 )[c]{$a_\mu^{(b-R)}$:}
%
  \DashLine(   310, 40 )( 370, 40 ){4}
  \PhotonArc(  340, 40 )( 30, 0, 180 ){5}{8}
  \CArc(       340, 40 )( 30, 0, 180 )
  \Line( 270, 40 )( 310, 40 )
  \Line( 370, 40 )( 410, 40 )
  \Text( 390, 40 )[c]{\huge $\times$}
  \Text( 266, 40 )[r]{$\mu_R$}
  \Text( 414, 40 )[l]{$e_L$}
  \Text( 310, 31 )[c]{$Y_\mu$}
  \Text( 370, 31 )[c]{$Y_e$}
  \Text( 340, 83 )[c]{$\Higgsino_d$}
  \Text( 341.3, 40 )[c]{{\large $\otimes$}}
  \Text( 330, 28 )[c]{$\snumu$}
  \Text( 352, 28 )[c]{$\snue$}
  \Text( 230, 40 )[c]{$a_{\mu e \ph}^{(b-R)}$:}
\end{picture}
\begin{picture}(440,90)
%
  \DashLine(    80, 40 )( 140, 40 ){4}
  \PhotonArc(  110, 40 )( 30, 0, 180 ){5}{8}
  \CArc(       110, 40 )( 30, 0, 180 )
  \Line(  40, 40 )( 80, 40 )
  \Line( 140, 40 )( 180, 40 )
  \Text( 110, 70 )[c]{\huge $\times$}
  \Text(  36, 40 )[r]{$\mu_R$}
  \Text( 184, 40 )[l]{$\mu_L$}
  \Text(  80, 31 )[c]{$Y_\mu$}
  \Text( 140, 32 )[c]{$g_2$}
  \Text(  90, 80 )[c]{$\Higgsino_d$}
  \Text( 130, 80 )[c]{$\Wino$}
  \Text( 111, 28 )[c]{$\snumu$}
  \Text(   0, 40 )[c]{$a_\mu^{(c-R)}$:}
%
  \DashLine(   310, 40 )( 370, 40 ){4}
  \PhotonArc(  340, 40 )( 30, 0, 180 ){5}{8}
  \CArc(       340, 40 )( 30, 0, 180 )
  \Line( 270, 40 )( 310, 40 )
  \Line( 370, 40 )( 410, 40 )
  \Text( 340, 70 )[c]{\huge $\times$}
  \Text( 266, 40 )[r]{$\mu_R$}
  \Text( 414, 40 )[l]{$e_L$}
  \Text( 310, 31 )[c]{$Y_\mu$}
  \Text( 370, 32 )[c]{$g_2$}
  \Text( 320, 80 )[c]{$\Higgsino_d$}
  \Text( 360, 80 )[c]{$\Wino$}
  \Text( 341.3, 40 )[c]{{\large $\otimes$}}
  \Text( 330, 28 )[c]{$\snumu$}
  \Text( 352, 28 )[c]{$\snue$}
  \Text( 230, 40 )[c]{$a_{\mu e \ph}^{(c-R)}$:}
\end{picture}
\caption{Chargino-sneutrino contributions to $a_r$ that give rise 
to muon $g-2$ and $\mu \ra e \ph$ in the interaction 
eigenstate basis.  The photon (not shown) is emitted from the chargino.
The chirality flip is shown by the $\times$
on the fermion line, while the the lepton flavor violating mass insertion
is shown by the $\otimes$.}
\label{char-R-fig}
\end{figure}

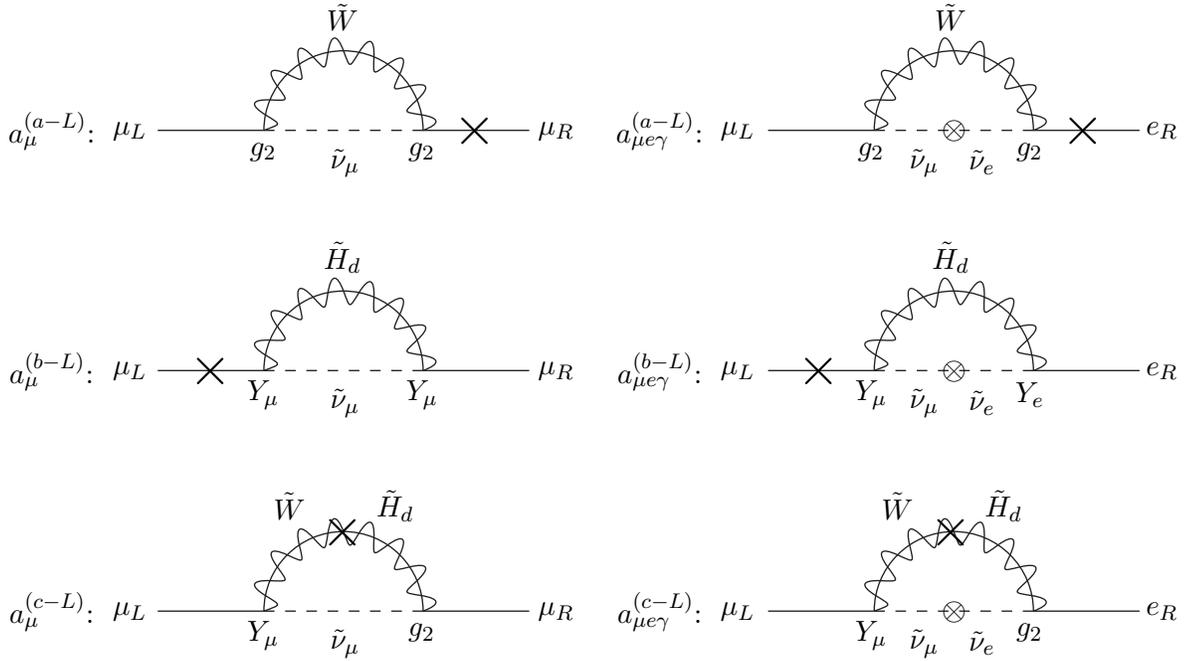
\begin{figure}[t]
\begin{picture}(440,90)
%
  \DashLine(    80, 40 )( 140, 40 ){4}
  \PhotonArc(  110, 40 )( 30, 0, 180 ){5}{8}
  \CArc(       110, 40 )( 30, 0, 180 )
  \Line(  40, 40 )( 80, 40 )
  \Line( 140, 40 )( 180, 40 )
  \Text( 160, 40 )[c]{\huge $\times$}
  \Text(  36, 40 )[r]{$\mu_L$}
  \Text( 184, 40 )[l]{$\mu_R$}
  \Text(  80, 32 )[c]{$g_2$}
  \Text( 140, 32 )[c]{$g_2$}
  \Text( 110, 83 )[c]{$\Wino$}
  \Text( 111, 28 )[c]{$\snumu$}
  \Text(   0, 40 )[c]{$a_\mu^{(a-L)}$:}
%
  \DashLine(   310, 40 )( 370, 40 ){4}
  \PhotonArc(  340, 40 )( 30, 0, 180 ){5}{8}
  \CArc(       340, 40 )( 30, 0, 180 )
  \Line( 270, 40 )( 310, 40 )
  \Line( 370, 40 )( 410, 40 )
  \Text( 390, 40 )[c]{\huge $\times$}
  \Text( 266, 40 )[r]{$\mu_L$}
  \Text( 414, 40 )[l]{$e_R$}
  \Text( 310, 32 )[c]{$g_2$}
  \Text( 370, 32 )[c]{$g_2$}
  \Text( 340, 83 )[c]{$\Wino$}
  \Text( 341.3, 40 )[c]{{\large $\otimes$}}
  \Text( 330, 28 )[c]{$\snumu$}
  \Text( 352, 28 )[c]{$\snue$}
  \Text( 230, 40 )[c]{$a_{\mu e \ph}^{(a-L)}$:}
\end{picture}
\begin{picture}(440,90)
%
  \DashLine(    80, 40 )( 140, 40 ){4}
  \PhotonArc(  110, 40 )( 30, 0, 180 ){5}{8}
  \CArc(       110, 40 )( 30, 0, 180 )
  \Line(  40, 40 )( 80, 40 )
  \Line( 140, 40 )( 180, 40 )
  \Text(  60, 40 )[c]{\huge $\times$}
  \Text(  36, 40 )[r]{$\mu_L$}
  \Text( 184, 40 )[l]{$\mu_R$}
  \Text(  80, 31 )[c]{$Y_\mu$}
  \Text( 140, 31 )[c]{$Y_\mu$}
  \Text( 110, 83 )[c]{$\Higgsino_d$}
  \Text( 111, 28 )[c]{$\snumu$}
  \Text(   0, 40 )[c]{$a_\mu^{(b-L)}$:}
%
  \DashLine(   310, 40 )( 370, 40 ){4}
  \PhotonArc(  340, 40 )( 30, 0, 180 ){5}{8}
  \CArc(       340, 40 )( 30, 0, 180 )
  \Line( 270, 40 )( 310, 40 )
  \Line( 370, 40 )( 410, 40 )
  \Text( 290, 40 )[c]{\huge $\times$}
  \Text( 266, 40 )[r]{$\mu_L$}
  \Text( 414, 40 )[l]{$e_R$}
  \Text( 310, 31 )[c]{$Y_\mu$}
  \Text( 370, 31 )[c]{$Y_e$}
  \Text( 340, 83 )[c]{$\Higgsino_d$}
  \Text( 341.3, 40 )[c]{{\large $\otimes$}}
  \Text( 330, 28 )[c]{$\snumu$}
  \Text( 352, 28 )[c]{$\snue$}
  \Text( 230, 40 )[c]{$a_{\mu e \ph}^{(b-L)}$:}
\end{picture}
\begin{picture}(440,90)
%
  \DashLine(    80, 40 )( 140, 40 ){4}
  \PhotonArc(  110, 40 )( 30, 0, 180 ){5}{8}
  \CArc(       110, 40 )( 30, 0, 180 )
  \Line(  40, 40 )( 80, 40 )
  \Line( 140, 40 )( 180, 40 )
  \Text( 110, 70 )[c]{\huge $\times$}
  \Text(  36, 40 )[r]{$\mu_L$}
  \Text( 184, 40 )[l]{$\mu_R$}
  \Text(  80, 31 )[c]{$Y_\mu$}
  \Text( 140, 32 )[c]{$g_2$}
  \Text(  90, 80 )[c]{$\Wino$}
  \Text( 130, 80 )[c]{$\Higgsino_d$}
  \Text( 111, 28 )[c]{$\snumu$}
  \Text(   0, 40 )[c]{$a_\mu^{(c-L)}$:}
%
  \DashLine(   310, 40 )( 370, 40 ){4}
  \PhotonArc(  340, 40 )( 30, 0, 180 ){5}{8}
  \CArc(       340, 40 )( 30, 0, 180 )
  \Line( 270, 40 )( 310, 40 )
  \Line( 370, 40 )( 410, 40 )
  \Text( 340, 70 )[c]{\huge $\times$}
  \Text( 266, 40 )[r]{$\mu_L$}
  \Text( 414, 40 )[l]{$e_R$}
  \Text( 310, 31 )[c]{$Y_\mu$}
  \Text( 370, 32 )[c]{$g_2$}
  \Text( 320, 80 )[c]{$\Wino$}
  \Text( 360, 80 )[c]{$\Higgsino_d$}
  \Text( 341.3, 40 )[c]{{\large $\otimes$}}
  \Text( 330, 28 )[c]{$\snumu$}
  \Text( 352, 28 )[c]{$\snue$}
  \Text( 230, 40 )[c]{$a_{\mu e \ph}^{(c-L)}$:}
\end{picture}
\caption{Similar to Fig.~\ref{char-L-fig}, the
chargino-sneutrino contributions to $a_l$ that give rise 
to muon $g-2$ and $\mu \ra e \ph$ in the interaction 
eigenstate basis.}
\label{char-L-fig}
\end{figure}

The complete expressions for the supersymmetric contributions to the  
muon anomalous magnetic moment have appeared in
Refs.~\cite{Moroi,CGW,MW}.
In general, muon (g - 2) arises from chargino-sneutrino and
neutralino-smuon graphs, with contributions differing
due to: the gaugino in the loop, the sfermion in the loop,
and location of the chirality flip(s).  In
Figs.~\ref{char-R-fig}--\ref{neut-L-LR-fig}
we show the supersymmetric diagrams that give rise to muon (g - 2). 
In order to obtain the corresponding $\mu \ra e \gamma$ graph 
for each of these diagrams we simply insert a flavor
violating soft mass ${m_{e \mu}}^2$ along the slepton line 
(other discussions of lepton flavor violation can be found in
e.g., Refs.~\cite{earlylfv,Sutter,GGMS}).
The resulting graph then gives an amplitude for $\mu \ra e \gamma$
that is related in a straightforward way to the original amplitude for
muon (g - 2) and also to the mass insertion ${m_{e \mu}}^2$. We can then
use the upper limit on the branching ratio for the process 
$\mu \ra e \gamma$ in the literature to obtain a bound on the flavor 
violating soft
mass ${m_{e \mu}}^2$. This bound will of course be crucially dependent on
the supersymmetric amplitude for muon (g - 2), which is an experimental
input.

Naively one might think that these flavor violating diagrams may be
heavily suppressed compared to the (g - 2) diagrams by a ratio of the
electron mass to the muon mass or more since each (g - 2) graph depends
explicitly on the (muon) flavor through one or more powers of the fermion
mass (or the fermion Yukawa coupling), resulting in a very weak bound.
However the detailed analysis we perform in subsequent sections reveals
that this is \emph{not} the case. We show that the supersymmetric 
contribution to (g -2) is dominated by graphs whose flavor violating 
analogues have no suppression factors, resulting in a very stringent 
bound.

Once constraints are established on the flavor violating mass
mixing, they can be immediately applied to supersymmetry breaking
models.  Generally the constraints are most severe \cite{GGMS} 
in models that communicate supersymmetry breaking from a hidden sector 
through gravitational interactions \cite{supergravity}.  In this framework 
the effective size of soft SUSY breaking is given by Planck suppressed
operators such as
\begin{eqnarray}
\int d^4 \theta \frac{S^\dag S}{M_{\Pl}^2} L_i^\dag L_j
\end{eqnarray}
where $S$ and $S^\dag$ are hidden sector fields, while $L_i$   
and $L_j$ are MSSM chiral multiplets of generation $i$ and $j$.
The Planck suppressed operators need not respect global flavor  
symmetries, and so $i$ and $j$ can be different.
When the hidden sector fields acquire a SUSY breaking $F$-term  
the soft (mass)$^2$ generated is
\begin{eqnarray}
\frac{|F_S|^2}{M_{\Pl}^2}
\end{eqnarray}
for all entries of the mass matrix in flavor space.
Instead, current bounds on both quark and lepton flavor
violating processes require either that the mass matrix
is nearly diagonal, or that it is ``aligned'' to the Yukawa
couplings.  This is the supersymmetric flavor problem.

One solution is to push at least the first and second generation
scalar masses to be very heavy (order tens of TeV or greater)
\cite{2-1-models}, but
this is inconsistent with generating a large supersymmetric
contribution of (g - 2).  Other methods of generating nearly flavor
blind masses involve the gauge and gaugino fields in some
nontrivial way.  Gauge mediation \cite{gaugemediation} 
postulates that the dominant   
SUSY breaking is communicated to the MSSM through gauge interactions
with a heavy messenger sector charged under the SM gauge group.
Anomaly mediation \cite{anomalymediation} and 
gaugino mediation \cite{gauginomediation} postulate physically separating
the hidden sector and the MSSM across a small extra dimension.
Soft masses are generated for MSSM fields through either tree-level
interactions (gaugino mediation) or at higher orders via the
trace anomaly (anomaly mediation).  In either case, the result is
a nearly flavor blind soft mass spectrum.

Why then is SUSY breaking induced flavor violation important?  
We view
this as means to test these mechanisms by accessing the flavor-nondiagonal
structure.  For instance, gravity mediation does contribute at a
suppressed level to the soft SUSY breaking masses in gauge mediation,
and so constraints on the size of flavor violating soft masses
can be translated into upper bounds on the scale of supersymmetry
breaking.  Also, in both anomaly mediation and gaugino mediation there are
exponentially suppressed flavor violating contributions resulting
from the small wavefunction overlap on the visible brane of SUSY
breaking fields localized on the hidden brane.  The size of this
suppressed contribution is
\begin{eqnarray}
e^{-M L} \frac{|F|^2}{M^2}
\end{eqnarray}
where $M = (L^{-1} M_{\Pl}^2)^{1/3}$ is the effective Planck
scale of the theory.  Constraining the the size of flavor
violation therefore restricts the size of the extra dimension,
requiring it to be roughly an order of magnitude \emph{larger} than
the effective Planck length.  Conversely, observing flavor
violation in this framework would allow an estimate of the size 
of the extra dimension.

Why is \emph{lepton} flavor violation important? 
The large body of evidence for
neutrino oscillations shows that lepton number cannot be an exact symmetry
of nature, and hence there is flavor physics that is outside the CKM
matrix.  Probing the sensitivity of the supersymmetry breaking sector 
to this flavor physics may provide insight into the scale and the nature 
of new flavor structure.  It would be interesting to see how 
constraints on SUSY induced lepton flavor
violation relate, if at all, to lepton violation through
neutrino masses (e.g., see \cite{neutrino-lfv}).  
We leave this for further study.

This paper is organized as follows.  In Sec.~\ref{char-sec}
we discuss the chargino-sneutrino contributions to (g - 2),
and in Sec.~\ref{char-lfv-sec} we calculate the chargino-sneutrino
contribution to the lepton flavor violating process $\mu \ra e\ph$.  
We then use the current experimental bounds on the branching ratio 
into $\mu\ra e\ph$ to establish a bound on the sneutrino (left-left) 
$e \leftrightarrow \mu$ flavor mixing mass.
In Sec.~\ref{neut-sec} we carry out the same analysis for
the neutralino-slepton contributions, and obtain similar
bounds on the left-left and right-right slepton flavor mixing masses
between $e \leftrightarrow \mu$.
In Sec.~\ref{tau-sec} we calculate the bounds on the flavor
violating process $\tau \ra \mu\ph$.  We find significantly
weaker constraints on the $\mu \leftrightarrow \tau$ flavor 
mixing mass as compared with the $e \leftrightarrow \mu$ flavor 
mixing mass.  Finally, in Sec.~\ref{conclusions-sec} we present
our conclusions.

\section{Chargino-sneutrino contributions}
\label{char-sec}

We begin by calculating the contributions to flavor violation 
resulting from sneutrino mixing.  These are the analog processes to 
the sneutrino contributions to (g - 2).  It is rather instructive 
to do this case analytically in detail, and so we restrict 
to considering $\snue$ -- $\snumu$ mixing only.

Suppose a small flavor violating (mass)$^2$ $m_{12}^2 \ll m_{\snue,\snumu}^2$ 
is added to the MSSM
\begin{eqnarray}
{\cal L} &=& m_{\snue}^2 \snue^*\snue + m_{\snumu}^2 \snumu^*\snumu 
             + m_{12}^2 \snue^* \snumu + h.c. \; ,
\end{eqnarray}
giving a simple $2\times2$ mass matrix for the sneutrinos,
\begin{eqnarray}
{\cal M}_{\snu}^2 &=& 
    \left( \begin{array}{cc} m_{\snue}^2 & m_{12}^2 \\ 
                             m_{12}^2 & m_{\snumu}^2 \end{array} \right) \; .
\end{eqnarray}
Note that we have absorbed the electroweak $D$-terms into our
definition of $m_{\sll}^2$ for any $\ell = \snue,\snumu,\se,\smu$.
Following the usual procedure we diagonalize this matrix, defining 
the mass eigenstates $\snuonetwo$ as
\begin{eqnarray}
\left( \begin{array}{c} \snuone \\ \snutwo \end{array} \right) &=&
\left( \begin{array}{cc} \cv & \sv \\ -\sv & \cv \end{array} \right)
\left( \begin{array}{c} \snue \\ \snumu \end{array} \right) 
\label{snu-mixing-eq}
\end{eqnarray}
ordered such that $\snuone$ is the sneutrino that is majority $\snue$ 
(and $\snutwo$ is majority $\snumu$).  The sneutrino mass mixing angle 
is given by
\begin{eqnarray}
\stwov &=& \frac{2 m_{12}^2}{\sqrt{\left( m_{\snumu}^2 - m_{\snue}^2 \right)^2
    + 4 m_{12}^2}}
\end{eqnarray}

The interactions of the sneutrino mass eigenstates with charginos 
can be obtained by inserting the mass eigenstates 
into the interaction Lagrangian \cite{GunionHaber},
\begin{eqnarray}
{\cal L} &=& \sum_k 
    \Big[ 
    \overline{e} \left( -g P_R V_{k1} + P_L Y_e U^*_{k2} \right) 
        \tilde{\chi}_k^c \left( \snuone \cv - \snutwo \sv \right)
\nonumber \\ & &{} \qquad
  + \overline{\mu} \left( -g P_R V_{k1} + P_L Y_\mu U^*_{k2} \right) 
        \tilde{\chi}_k^c \left( \snuone \sv + \snutwo \cv \right)
    \Big] + h.c. \; .
\label{mix-lagrangian-eq}
\end{eqnarray}
We now calculate the chargino-sneutrino contribution to (g - 2) 
in terms of the mass eigenstates.  There are three distinct
contributions that can be identified by the gaugino running
in the loop and the location of the chirality flip:  (a) pure gauge, 
(b) pure Higgsino, and (c) mixed, as shown in 
Figs.~\ref{char-R-fig},\ref{char-L-fig}. {\footnote{These diagrams are in
the
interaction basis and are therefore schematic. However this is merely for
purposes of clarity and does not affect our conclusions.}}
 The total
chargino-sneutrino contribution is simply
\begin{eqnarray}
a_\mu &=& a_\mu^{(a)} + a_\mu^{(b)} + a_\mu^{(c)} \; ,
\end{eqnarray}
where the individual diagrams give
\begin{eqnarray}
\frac{16\pi^2}{m_\mu} a_\mu^{(a)} &=& \frac{m_\mu}{12 m_{\Cpm_k}^2}
    g_2^2 |V_{k1}|^2 \left[ \twosv x_{k1} F_1^C(x_{k1}) 
    + \twocv x_{k2} F_1^C(x_{k2}) \right] \label{g-2-a-eq} \\
\frac{16\pi^2}{m_\mu} a_\mu^{(b)} &=& \frac{m_\mu}{12 m_{\Cpm_k}^2}
    Y_\mu^2 |U_{k2}|^2 \left[ \twosv x_{k1} F_1^C(x_{k1}) 
    + \twocv x_{k2} F_1^C(x_{k2}) \right] \label{g-2-b-eq} \\
\frac{16\pi^2}{m_\mu} a_\mu^{(c)} &=& -\frac{2}{3 m_{\Cpm_k}}
    g_2 Y_\mu {\rm Re}[V_{k1} U_{k2}] \left[ \twosv x_{k1} F_2^C(x_{k1}) 
    + \twocv x_{k2} F_2^C(x_{k2}) \right] \label{g-2-c-eq} \; ,
\end{eqnarray}
where the sum over $k=1,2$ for the two charginos is implicitly understood.
Here $Y_\mu$ is the muon Yukawa coupling, $g_2$ is the SU(2)$_L$ coupling,
$x_{1k,2k} \equiv m_{\chi_k^\pm}^2/m_{\snuonetwo}^2$, and 
$U_{ij}, V_{ij}$ are the chargino mixing matrices
in the  $(\Wino^\pm,\Higgsino^\pm)$ basis.
In the Appendix we provide the one-loop kinematical functions $F(x)$,
which are defined identically to Ref.~\cite{MW}.
In the limit of no flavor violation, $m_{12}^2 \ra 0$,
this result agrees with previous calculations \cite{CGW,MW}.

Given that the size of flavor mixing mass is small, 
$m_{12}^2 \ll m_{\snue}^2, m_{\snumu}^2$, there are two limiting 
cases:  $m_{12}^2 \ll |m_{\snumu}^2 - m_{\snue}^2|$,
and $m_{12}^2 \gg |m_{\snumu}^2 - m_{\snue}^2|$.  
In the first case, the mixing angle $\thetav \ll 1$, so that $\cv \simeq 1$ 
while $\sv \ll 1$.  The mass eigenstates are therefore nearly identical
to the interaction eigenstates $\snutwo \simeq \snumu$ and 
$\snuone \simeq \snue$, so that Eqs.~(\ref{g-2-a-eq})--(\ref{g-2-c-eq})
trivially reduces to the interaction eigenstate result.  
In the second case, the mixing angle is maximal $\thetav \simeq \pi/4$, 
so both sneutrinos contribute about equally but suppressed by
a factor of $\twosv \simeq \twocv \simeq 1/2$.  The sum of course
gives nearly the same result as obtained without flavor mixing.  
Thus, our formalism gives the expected result that small flavor mixing 
in the sneutrino masses does not affect the prediction of (g - 2) 
in supersymmetry.  We can express this result as
\begin{eqnarray}
\twosv x_{k1} F_{1,2}^C(x_{k1}) + \twocv x_{k2} F_{1,2}^C(x_{k2})
   &\simeq& x_{k,\mu} F_{1,2}^C(x_{k,\mu}) \; ,
\end{eqnarray}
where $x_{k,\mu} \equiv m_{\chi_k^\pm}^2/m_{\snumu}^2$.

What is the relative size of each of these diagrams?
Here the advantage of working in the interaction eigenstate
basis is apparent.  In the limit $M_2 \ll \mu$, the gaugino diagram 
dominates, and the others can be ignored.  When $M_2 \sim \mu$, there is 
large gaugino-Higgsino mixing $|V_{k1}| \sim |U_{k2}|$, and so the relative 
size of the diagrams is governed by the couplings:
\begin{eqnarray}
\left|a_\mu^{(a)}\right| \, 
  : \, \left|a_\mu^{(b)}\right| \, 
  : \, \left|a_\mu^{(c)}\right| &=&
\left|\frac{g_2 m_\mu}{8 Y_\mu m_{\Cpm}}\right| \, 
  : \, \left|\frac{Y_\mu m_\mu}{8 g_2 m_{\Cpm}}\right| \, 
  : \, 1 \; .
\end{eqnarray}
In this case clearly diagram (b) is highly suppressed relative to diagram
(a) or (c).  The relative competition between diagram (a) and
(c) depends on $\tan\beta$ and the mass of the chargino.
Finally, the limit $\mu \ll M_2$ is slightly subtle.  Fortunately
it is straightforward to show that in this limit 
$|V_{k1} U_{k2}| \simeq \sqrt{2} M_W/M_2$, and so the ratio 
\begin{eqnarray}
\left|\frac{a_\mu^{(c)}}{a_\mu^{(b)}}\right| &\sim& 
    \frac{g_2 \mu M_W}{Y_\mu m_\mu M_2} \; .
\end{eqnarray}
This ratio is much greater than one even for $\mu/M_2 = (m_\mu/M_W)^{_1}
\sim 10^3$.
Diagram (c) therefore dominates in the light Higgsino case.

Hence, the overwhelmingly dominant chargino-sneutrino contribution 
to (g - 2) arises from diagrams (a) and (c).  Hereafter, we ignore diagram (b).

\subsection{Flavor violating chargino-sneutrino graphs}
\label{char-lfv-sec}

The chargino-sneutrino contribution to lepton flavor violation
is related to the muon (g - 2) graph by a simple replacement of 
the outgoing muon with an electron. The (g - 2) graphs can be written
in pairs, with the same particles in the loop but the ingoing and
outgoing muons having different chiralities. These graphs have the same
amplitude if there are no phases.
For $\mu\ra e\ph$, however, there are two distinct sets of diagrams 
for the left-handed and right-handed incoming muons, leading to 
distinct contributions to the amplitude for $a_l$ and $a_r$.
We find that the contribution to $a_l$ is suppressed by 
at least one additional power of the electron mass or 
electron Yukawa coupling, as so can be neglected.
The contribution to $a_r$ can be split into the
same three contributions as we did above for (g - 2).  We obtain
\begin{eqnarray}
\frac{16\pi^2}{m_\mu} a_{\mu e \ph}^{(a)} &=& 
    \frac{m_\mu}{24 m_{\Cpm_k}^2}
    g_2^2 |V_{k1}|^2 \stwov 
    \left[ x_{k1} F_1^C(x_{k1}) - x_{k2} F_1^C(x_{k2}) \right] \\
\frac{16\pi^2}{m_\mu} a_{\mu e \ph}^{(b)} &=& 
    \frac{m_\mu}{24 m_{\Cpm_k}^2}
    Y_\mu Y_e |U_{k2}|^2 \stwov 
    \left[ x_{k1} F_1^C(x_{k1}) - x_{k2} F_1^C(x_{k2}) \right] \\
\frac{16\pi^2}{m_\mu} a_{\mu e \ph}^{(c)} &=& -\frac{1}{3 m_{\Cpm_k}}
    g_2 Y_\mu {\rm Re}[V_{k1} U_{k2}] \stwov 
    \left[ x_{k1} F_2^C(x_{k1}) - x_{k2} F_2^C(x_{k2}) \right] \; .
\end{eqnarray}
Following the arguments we made for (g - 2), it is an excellent
approximation to neglect diagram (b).  The relationship between
the amplitudes for (g - 2) and $\mu \ra e\ph$ can be seen by
writing the ratio of diagrams (for fixed $k$, not summed over),
\begin{eqnarray}
\frac{a_{\mu e \ph}^{(i)}}{a_\mu^{(i)}} &=& \frac{1}{2} 
  \frac{\stwov \left( x_{k1} F^C(x_{k1}) - x_{k2} F^C(x_{k2}) 
        \right)}{x_{k,\mu} F^C(x_{k,\mu})} \; .
\label{char-ratio-eq}
\end{eqnarray}
Here $F_1(x) = F_1^C(x)$ for $(i)=(a)$ and $F_1(x) = F_2^C(x)$ 
for $(i)=(c)$.  Notice that the couplings and chargino mixing
angle drops out.  The expression can be evaluated (exactly)
in the $2\times2$ mixing case. However before doing this we first
obtain a qualitative understanding of this ratio of amplitudes.
    
From the expression above it is clear that unless
\begin{equation}
\label{eq:limit}
\frac{x_{k1} F^C(x_{k1})}{x_{k2} F^C(x_{k2})} - 1 \ll 1
\end{equation}
the ratio of amplitudes
\begin{equation}
\frac{a_{\mu e \ph}^{(i)}}{a_\mu^{(i)}} \simeq \frac{1}{2} \stwov
\simeq \frac{m_{12}^2}{\mbox{Max}[m_{\snumu}^2,m_{\snue}^2]}
\end{equation}

Now the approximation (\ref{eq:limit}) that leads to the equation
above clearly breaks down in the limit
$x_{k1} \ra x_{k2}$. A careful examination of the structure of the
functions $F^C(x)$ also shows that (\ref{eq:limit}) is not satisfied when
both $x_{k1} \gg 1$ and $x_{k2} \gg 1$ even if $x_{k1}$ and $x_{k2}$ are
very different in magnitude. Since the functions $F^C(x)$ are monotonic   
apart from these regions the equation (\ref{eq:limit}) is satisfied. We
therefore examine these two limits in further detail.

We first consider the case of $x_{k1} \gg 1$, $x_{k2} \gg 1$. Then by
studying the asymptotic behavior of $F^C(x)$ it follows that
\begin{equation}
\frac{a_{\mu e \ph}^{(i)}}{a_\mu^{(i)}} \simeq
\frac{m_{12}^2}{m_{\Cpm_k}^2}
\end{equation}  
This suggests that Eq.~(\ref{eq:limit}) can be generalized to
\begin{eqnarray}
\frac{a_{\mu e \ph}^{(i)}}{a_\mu^{(i)}} &\simeq&
    \frac{m_{12}^2}{\mbox{Max}[m_{\snumu}^2,m_{\snue}^2,m_{\Cpm_k}^2]}
    \label{eq:limit1}
\end{eqnarray}

We now consider $x_{k1} \ra x_{k2}$. For simplicity we examine  
the limits $|m_{\snumu}^2 - m_{\snue}^2| \gg m_{12}^2$ and
$|m_{\snumu}^2 - m_{\snue}^2| \ll m_{12}^2$ separately.
For $|m_{\snumu}^2 - m_{\snue}^2| \gg m_{12}^2$ we have
\begin{eqnarray}
\stwov &\simeq& \frac{2 m_{12}^2}{|m_{\snumu}^2 - m_{\snue}^2|}     
\end{eqnarray}
and
\begin{eqnarray} 
\frac{ x_{k1} F^C(x_{k1}) - x_{k2} F^C(x_{k2})}{x_{k,\mu} F^C(x_{k,\mu})} 
    &\simeq& \frac{m_{\snumu}^2 - m_{\snue}^2}{m_{\snumu}^2} 
    \qquad x_{k} \lsim 1 \\
\frac{x_{k1} F^C(x_{k1}) - x_{k2} F^C(x_{k2})}{x_{k,\mu} F^C(x_{k,\mu})} 
    &\simeq& \frac{m_{\snumu}^2 - m_{\snue}^2}{m_{\Cpm_k}^2} 
    \qquad x_{k} \gg 1
\end{eqnarray}
up to numerical factors of order one.
Combining these equations we find (\ref{eq:limit1}) is in fact reproduced
even in this limit. Now consider the case $x_{k1} \ra x_{k2}$
with $\arrowvert m_{\snumu}^2 - m_{\snue}^2 \arrowvert \ll
m_{12}^2$. Then $\stwov \simeq 1$ while
\begin{eqnarray}
\frac{ x_{k1} F^C(x_{k1}) - x_{k2} F^C(x_{k2})
        }{x_{k,\mu} F^C(x_{k,\mu})} \simeq& \frac{m_{12}^2}
{m_{\snumu}^2} & \qquad x_{k} \lsim 1\\
\frac{ x_{k1} F^C(x_{k1}) - x_{k2} F^C(x_{k2})
        }{x_{k,\mu} F^C(x_{k,\mu})} \simeq& \frac{
m_{12}^2}{m_{\Cpm_k}^2} & \qquad x_{k} \gg 1  
\end{eqnarray}

Hence once again we find \ref{eq:limit1} holds. It is not difficult to
verify that this remains true in the intermediate region
 $\left[\arrowvert m_{\snumu}^2 - m_{\snue}^2 \arrowvert \right] \approx
m_{12}^2$. From this simple analysis we therefore conclude that the
amplitudes for $\mu \ra e \gamma$ and (g - 2) are simply related by
\begin{eqnarray}
\frac{a_{\mu e \ph}^{(i)}}{a_\mu^{(i)}} &\simeq& 
    \frac{m_{12}^2}{\mbox{Max}[m_{\snumu}^2,m_{\snue}^2,m_{\Cpm_k}^2]} 
\end{eqnarray}
It is not difficult to understand the origin of this
result. Given the correspondence between diagrams then
elementary dimensional considerations indicate that the ratio
of the two amplitudes is approximately given by $\frac{m_{12}^2}{M^2}$
where $M^2$ is a heavy scale. The most conservative assumption is that
$M^2$ is the heaviest scale in the problem, the mass of the heaviest
particle, which immediately yields
Eq.~(\ref{eq:limit1}).

We now perform a more careful analysis of the problem.  
The simplest piece of Eq.~(\ref{char-ratio-eq}) to evaluate
is the denominator.  Since we are assuming no relations among
the soft supersymmetry breaking masses, let's consider the three 
possible limits:  (1) $x_{k,\mu} \gg 1$, (2) $x_{k,\mu} \sim 1$,
and (3) $x_{k,\mu} \ll 1$, corresponding to the chargino mass being
much greater than, roughly equal to, or much smaller than the muon 
sneutrino mass, respectively.  The limits of the one-loop functions 
are given in the Appendix, and so we simply state the result here:
\begin{eqnarray}
  &   & \frac{2}{m_{\Cpm_k}^2} \qquad {\rm for} \; x_{k,\mu} \gg 1 \\
\frac{1}{m_{\Cpm_k}^2} x_{k,\mu} F_1^C(x_{k,\mu}) 
  & = & \frac{2}{m_{\snumu}^2} \qquad {\rm for} \; x_{k,\mu} \sim 1 \\
  &   & \frac{4}{m_{\snumu}^2} \qquad {\rm for} \; x_{k,\mu} \ll 1 \; ,
\end{eqnarray}
which can be written very roughly as
\begin{eqnarray}
\frac{1}{m_{\Cpm_k}^2} x_{k,\mu} F_1^C(x_{k,\mu}) 
  &\sim& \frac{1}{\mbox{Max}[m_{\Cpm_k}^2, m_{\snumu}^2]}
\end{eqnarray}
for any $x$, dropping overall factors of $2$.  The same expression 
can be found for the other one-loop functions with one exception.\footnote{The 
exception is the small $x$ limit of $x F_2^C(x)$, which behaves 
as $-x \ln x$.}

The numerator of Eq.~(\ref{char-ratio-eq}) can also be evaluated
for analogous limits.  Since both sneutrino masses are
in the expression there are nine distinct cases depending on
the relative hierarchy of $m_{\snuone}$, $m_{\snutwo}$, and $m_{\Cpm_k}$.  
The ratio can be evaluated straightforwardly in all of these cases.
We present the results in Table~\ref{char-table}.
\begin{table}[t]
\renewcommand{\arraystretch}{1.5}\small\normalsize
\setlength{\tabcolsep}{1cm}
\begin{center}
\begin{tabular}{ccc} \hline\hline
case & $a_{\mu e \ph}^{(a)}/a_\mu^{(a)}$ & $a_{\mu e \ph}^{(c)}/a_\mu^{(c)}$ 
   \\ \hline
$x_{k,1},x_{k,2} \gg 1$           & $2m_{12}^2/m_{\Cpm_k}^2$ & 
    $ m_{12}^2/m_{\Cpm_k}^2$ \\
$x_{k,1},x_{k,2} \ll 1$           & $ m_{12}^2/m_{\snue}^2$ & 
    $ m_{12}^2/m_{\snue}^2$ \\
$x_{k,1} \ll 1 \ll x_{k,2}$       & $ m_{12}^2/m_{\snue}^2$ & 
    $ m_{12}^2/m_{\snue}^2$ \\
$x_{k,2} \ll 1 \ll x_{k,1}$       & $ m_{12}^2/m_{\Cpm_k}^2$ & 
    $ \textfrac{1}{2}m_{12}^2/(m_{\Cpm_k}^2 \ln 
m_{\snumu}^2/m_{\Cpm_k}^2)$ \\
$x_{k,1} \sim x_{k,2} \sim 1$     & $\textfrac{2}{5} m_{12}^2/m_{\snue}^2$
& 
    $\textfrac{1}{4} m_{12}^2/m_{\snue}^2$ \\
$x_{k,1} \sim 1$, $x_{k,2} \gg 1$ & $\textfrac{1}{2} m_{12}^2/m_{\snue}^2$
& 
    $\textfrac{1}{3} m_{12}^2/m_{\snue}^2$ \\
$x_{k,1} \sim 1$, $x_{k,2} \ll 1$ & $\textfrac{1}{4}
m_{12}^2/m_{\Cpm_k}^2$ & 
    $\textfrac{1}{3} m_{12}^2/(m_{\Cpm_k}^2 \ln
m_{\snumu}^2/m_{\Cpm_k}^2)$ \\
$x_{k,2} \sim 1$, $x_{k,1} \gg 1$ & $ m_{12}^2/m_{\Cpm_k}^2$ & 
    $\textfrac{1}{2} m_{12}^2/m_{\Cpm_k}^2$ \\
$x_{k,2} \sim 1$, $x_{k,1} \ll 1$ & $ m_{12}^2/m_{\snue}^2$ & 
    $ m_{12}^2/m_{\snue}^2$ \\
\hline\hline
\end{tabular}
\end{center}
\caption{The ratio of the amplitude for $\mu \ra e\ph$ over (g - 2) for 
diagram (a) and (c), for a given chargino ($k$ fixed).}
\label{char-table}
\end{table}
The tiny electron mass effect was ignored ($m_e/m_\mu \ra 0$),
and for those cases with $x_{i,k} \sim 1$ we have freely interchanged 
$m_{\snu_i}$ with $m_{\Cpm_k}$.

The most important result is that each (g - 2) diagram has a
$\mu \ra e\ph$ counterpart that is proportional to 
\begin{eqnarray}
\delta_{12} = \frac{m_{12}^2}{\mbox{Max}[m_{\Cpm_k}^2,m_{\snue}^2,
m_{\snumu}^2]}
\end{eqnarray}
for \emph{any} choice of soft breaking parameters.  Right away we see 
that if a nonzero supersymmetric contribution to muon (g - 2) comes from 
mainly one diagram (with one chargino in the loop), then there is 
a prediction for the size of the lepton flavor violating process 
$\mu \ra e\ph$ that only depends on the size of the flavor mixing
(mass)$^2$ divided by the larger of the chargino mass and
the electron sneutrino mass.\footnote{In two of eighteen cases 
in Table~\ref{char-table} there is an additional logarithmic 
suppression of $1/\ln m_{\Cpm_k}^2/m_{\snumu}^2$.  However, the 
contribution to (g - 2) is suppressed as a power law 
$\propto m_{\Cpm_k}^2/m_{\snumu}^2$, and so this case is not 
relevant to our discussion.}
The prediction is
\begin{eqnarray}
a_{\mu e \ph} &\gsim& \frac{1}{4} a_\mu \delta_{12} \simeq a_r\; .
\end{eqnarray}
We take the low value of $a_{\mu e \ph}$ because we are interested in a
bound.
The width for $\mu \ra e\ph$ is easily obtained from 
the magnetic moment operator Eq.~(\ref{mag-moment-eq})
\begin{eqnarray}
\Gamma(\mu \ra e\ph) &=& \frac{m_\mu e^2}{64\pi} 
                         \left( |a_l|^2 + |a_r|^2 \right) \; .
\end{eqnarray}
and as we discussed above, we may ignore the $a_l$ contribution
since it is further suppressed by the electron mass.
The branching ratio is then
\begin{eqnarray}
\mbox{BR}(\mu\ra e\ph) &=& \frac{3 \pi^2 e^2}{G_F^2 m_\mu^4} |a_r|^2 \; .
\end{eqnarray}
which we can write as
\begin{eqnarray}
\mbox{BR}(\mu\ra e\ph) &\simeq& 2.0 \times 10^{-4}
  \left( \frac{a_\mu}{4.3 \times 10^{-9}} \right)^2 \delta_{12}^2
\end{eqnarray}
The current experimental bound is $\mbox{BR}(\mu\ra e\ph) < 1.2 \times 
10^{-11}$ \cite{MEGA}, which we can use to place a bound on the 
flavor mixing (mass)$^2$:
\begin{eqnarray}
\delta_{12} < 2.4 \times 10^{-4} 
    \left( \frac{\mbox{BR}(\mu \ra e\ph)}{1.2 \times 10^{-11}} \right)^{1/2} 
    \left( \frac{4.3 \times 10^{-9}}{a_\mu} \right)
\end{eqnarray}
This bound is as accurate as $\delta_{12}$ is known, i.e., 
to within about a factor of $2$.

\section{Neutralino-slepton contributions}
\label{neut-sec}

The second class of diagrams that contribute to muon (g - 2)
and lepton flavor violation are ones with neutralinos and 
charged sleptons in the loop.  We again restrict ourselves
to $\mu \leftrightarrow e$ flavor transitions only,
taking up other possibilities in Sec.~\ref{tau-sec}.
There are several important differences with the chargino-sneutrino
contributions:  
\begin{itemize}
\item there are four neutralinos $\Bino$, $\Wino^0$, $\Higgsino^0_d$,
      and $\Higgsino^0_u$ instead of two, and hence the mass matrix
      is four by four rather than two by two  
\item there is a pair of charged sleptons for each flavor
      $\se_{L,R}$ and $\smu_{L,R}$
\item left-right slepton mixing in addition to flavor mixing
\item flavor mixing can be between left-left, right-right, a 
      combination of both, and left-to-right or right-to-left
\end{itemize}

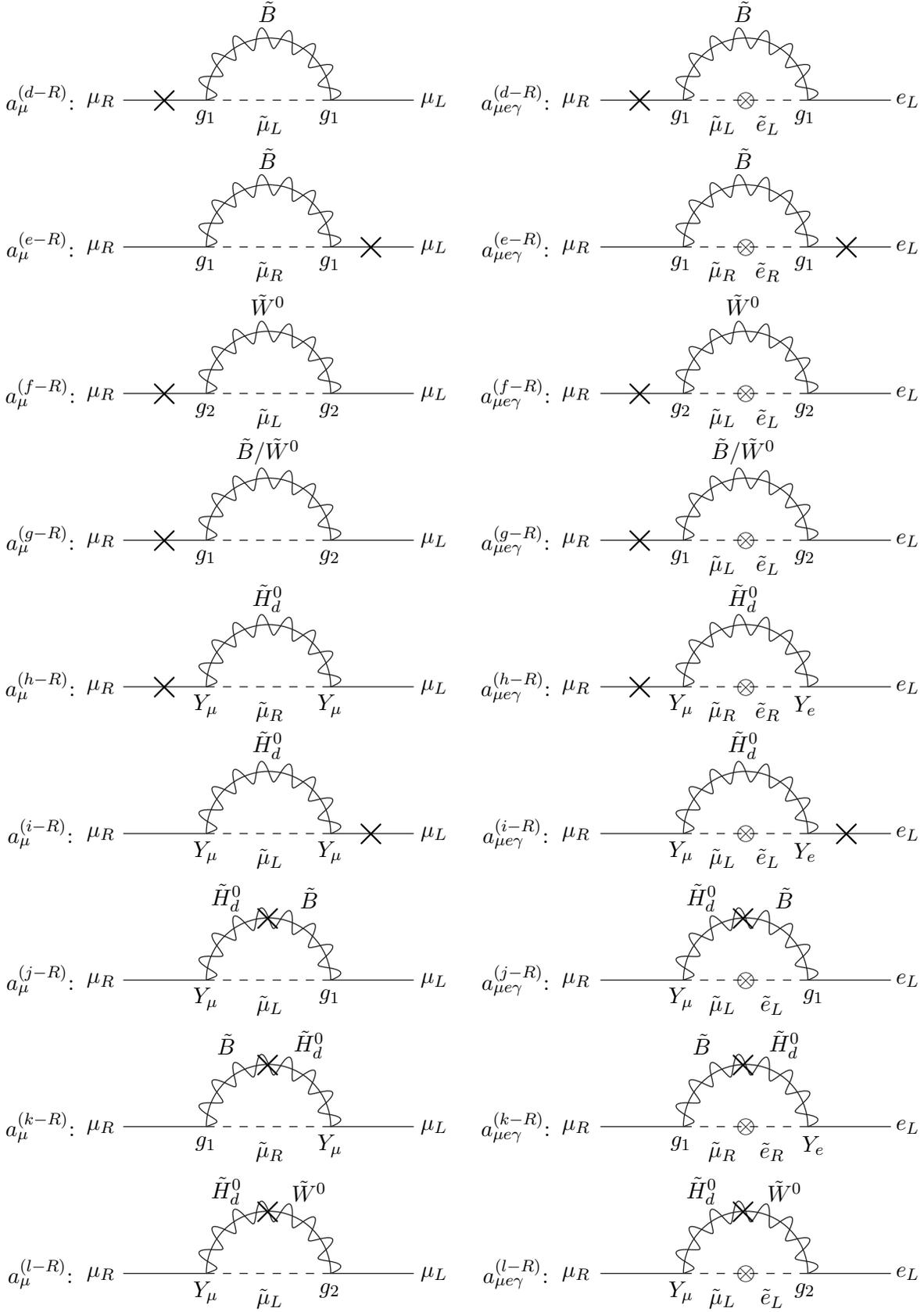
\begin{figure}[t]
\vspace*{-0.8cm}
\begin{picture}(440,70)
%
  \DashLine(    80, 40 )( 140, 40 ){4}
  \PhotonArc(  110, 40 )( 30, 0, 180 ){5}{8}
  \CArc(       110, 40 )( 30, 0, 180 )
  \Line(  40, 40 )( 80, 40 )
  \Line( 140, 40 )( 180, 40 )
  \Text(  60, 40 )[c]{\huge $\times$}
  \Text(  36, 40 )[r]{$\mu_R$}
  \Text( 184, 40 )[l]{$\mu_L$}
  \Text(  80, 32 )[c]{$g_1$}
  \Text( 140, 32 )[c]{$g_1$}
  \Text( 110, 83 )[c]{$\Bino$}
  \Text( 111, 28 )[c]{$\smu_L$}
  \Text(   0, 40 )[c]{$a_\mu^{(d-R)}$:}
%
  \DashLine(   310, 40 )( 370, 40 ){4}
  \PhotonArc(  340, 40 )( 30, 0, 180 ){5}{8}
  \CArc(       340, 40 )( 30, 0, 180 )
  \Line( 270, 40 )( 310, 40 )
  \Line( 370, 40 )( 410, 40 )
  \Text( 290, 40 )[c]{\huge $\times$}
  \Text( 266, 40 )[r]{$\mu_R$}
  \Text( 414, 40 )[l]{$e_L$}
  \Text( 310, 32 )[c]{$g_1$}
  \Text( 370, 32 )[c]{$g_1$}
  \Text( 340, 83 )[c]{$\Bino$}
  \Text( 341.3, 40 )[c]{{\large $\otimes$}}
  \Text( 330, 28 )[c]{$\smu_L$}
  \Text( 352, 28 )[c]{$\se_L$}
  \Text( 230, 40 )[c]{$a_{\mu e \ph}^{(d-R)}$:}
\end{picture}
\begin{picture}(440,70)
%
  \DashLine(    80, 40 )( 140, 40 ){4}
  \PhotonArc(  110, 40 )( 30, 0, 180 ){5}{8}
  \CArc(       110, 40 )( 30, 0, 180 )
  \Line(  40, 40 )( 80, 40 )
  \Line( 140, 40 )( 180, 40 )
  \Text( 160, 40 )[c]{\huge $\times$}
  \Text(  36, 40 )[r]{$\mu_R$}
  \Text( 184, 40 )[l]{$\mu_L$}
  \Text(  80, 32 )[c]{$g_1$}
  \Text( 140, 32 )[c]{$g_1$}
  \Text( 110, 83 )[c]{$\Bino$}
  \Text( 111, 28 )[c]{$\smu_R$}
  \Text(   0, 40 )[c]{$a_\mu^{(e-R)}$:}
%
  \DashLine(   310, 40 )( 370, 40 ){4}
  \PhotonArc(  340, 40 )( 30, 0, 180 ){5}{8}
  \CArc(       340, 40 )( 30, 0, 180 )
  \Line( 270, 40 )( 310, 40 )
  \Line( 370, 40 )( 410, 40 )
  \Text( 390, 40 )[c]{\huge $\times$}
  \Text( 266, 40 )[r]{$\mu_R$}
  \Text( 414, 40 )[l]{$e_L$}
  \Text( 310, 32 )[c]{$g_1$}
  \Text( 370, 32 )[c]{$g_1$}
  \Text( 340, 83 )[c]{$\Bino$}
  \Text( 341.3, 40 )[c]{{\large $\otimes$}}
  \Text( 330, 28 )[c]{$\smu_R$}
  \Text( 352, 28 )[c]{$\se_R$}
  \Text( 230, 40 )[c]{$a_{\mu e \ph}^{(e-R)}$:}
\end{picture}
\begin{picture}(440,70)
%
  \DashLine(    80, 40 )( 140, 40 ){4}
  \PhotonArc(  110, 40 )( 30, 0, 180 ){5}{8}
  \CArc(       110, 40 )( 30, 0, 180 )
  \Line(  40, 40 )( 80, 40 )
  \Line( 140, 40 )( 180, 40 )
  \Text(  60, 40 )[c]{\huge $\times$}
  \Text(  36, 40 )[r]{$\mu_R$}
  \Text( 184, 40 )[l]{$\mu_L$}
  \Text(  80, 32 )[c]{$g_2$}
  \Text( 140, 32 )[c]{$g_2$}
  \Text( 110, 83 )[c]{$\Wino^0$}
  \Text( 111, 28 )[c]{$\smu_L$}
  \Text(   0, 40 )[c]{$a_\mu^{(f-R)}$:}
%
  \DashLine(   310, 40 )( 370, 40 ){4}
  \PhotonArc(  340, 40 )( 30, 0, 180 ){5}{8}
  \CArc(       340, 40 )( 30, 0, 180 )
  \Line( 270, 40 )( 310, 40 )
  \Line( 370, 40 )( 410, 40 )
  \Text( 290, 40 )[c]{\huge $\times$}
  \Text( 266, 40 )[r]{$\mu_R$}
  \Text( 414, 40 )[l]{$e_L$}
  \Text( 310, 32 )[c]{$g_2$}
  \Text( 370, 32 )[c]{$g_2$}
  \Text( 340, 83 )[c]{$\Wino^0$}
  \Text( 341.3, 40 )[c]{{\large $\otimes$}}
  \Text( 330, 28 )[c]{$\smu_L$}
  \Text( 352, 28 )[c]{$\se_L$}
  \Text( 230, 40 )[c]{$a_{\mu e \ph}^{(f-R)}$:}
\end{picture}
\begin{picture}(440,70)
%
  \DashLine(    80, 40 )( 140, 40 ){4}
  \PhotonArc(  110, 40 )( 30, 0, 180 ){5}{8}
  \CArc(       110, 40 )( 30, 0, 180 )
  \Line(  40, 40 )( 80, 40 )
  \Line( 140, 40 )( 180, 40 )
  \Text(  60, 40 )[c]{\huge $\times$}
  \Text(  36, 40 )[r]{$\mu_R$}
  \Text( 184, 40 )[l]{$\mu_L$}
  \Text(  80, 32 )[c]{$g_1$}
  \Text( 140, 32 )[c]{$g_2$}
  \Text( 110, 83 )[c]{$\Bino/\Wino^0$}
  \Text(   0, 40 )[c]{$a_\mu^{(g-R)}$:}
%
  \DashLine(   310, 40 )( 370, 40 ){4}
  \PhotonArc(  340, 40 )( 30, 0, 180 ){5}{8}
  \CArc(       340, 40 )( 30, 0, 180 )
  \Line( 270, 40 )( 310, 40 )
  \Line( 370, 40 )( 410, 40 )
  \Text( 290, 40 )[c]{\huge $\times$}
  \Text( 266, 40 )[r]{$\mu_R$}
  \Text( 414, 40 )[l]{$e_L$}
  \Text( 310, 32 )[c]{$g_1$}
  \Text( 370, 32 )[c]{$g_2$}
  \Text( 340, 83 )[c]{$\Bino/\Wino^0$}
  \Text( 341.3, 40 )[c]{{\large $\otimes$}}
  \Text( 330, 28 )[c]{$\smu_L$}
  \Text( 352, 28 )[c]{$\se_L$}
  \Text( 230, 40 )[c]{$a_{\mu e \ph}^{(g-R)}$:}
\end{picture}
\begin{picture}(440,70)
%
  \DashLine(    80, 40 )( 140, 40 ){4}
  \PhotonArc(  110, 40 )( 30, 0, 180 ){5}{8}
  \CArc(       110, 40 )( 30, 0, 180 )
  \Line(  40, 40 )( 80, 40 )
  \Line( 140, 40 )( 180, 40 )
  \Text(  60, 40 )[c]{\huge $\times$}
  \Text(  36, 40 )[r]{$\mu_R$}
  \Text( 184, 40 )[l]{$\mu_L$}
  \Text(  80, 31 )[c]{$Y_\mu$}
  \Text( 140, 31 )[c]{$Y_\mu$}
  \Text( 110, 83 )[c]{$\Higgsino_d^0$}
  \Text( 111, 28 )[c]{$\smu_R$}
  \Text(   0, 40 )[c]{$a_\mu^{(h-R)}$:}
%
  \DashLine(   310, 40 )( 370, 40 ){4}
  \PhotonArc(  340, 40 )( 30, 0, 180 ){5}{8}
  \CArc(       340, 40 )( 30, 0, 180 )
  \Line( 270, 40 )( 310, 40 )
  \Line( 370, 40 )( 410, 40 )
  \Text( 290, 40 )[c]{\huge $\times$}
  \Text( 266, 40 )[r]{$\mu_R$}
  \Text( 414, 40 )[l]{$e_L$}
  \Text( 310, 31 )[c]{$Y_\mu$}
  \Text( 370, 31 )[c]{$Y_e$}
  \Text( 340, 83 )[c]{$\Higgsino_d^0$}
  \Text( 341.3, 40 )[c]{{\large $\otimes$}}
  \Text( 330, 28 )[c]{$\smu_R$}
  \Text( 352, 28 )[c]{$\se_R$}
  \Text( 230, 40 )[c]{$a_{\mu e \ph}^{(h-R)}$:}
\end{picture}
\begin{picture}(440,70)
%
  \DashLine(    80, 40 )( 140, 40 ){4}
  \PhotonArc(  110, 40 )( 30, 0, 180 ){5}{8}
  \CArc(       110, 40 )( 30, 0, 180 )
  \Line(  40, 40 )( 80, 40 )
  \Line( 140, 40 )( 180, 40 )
  \Text( 160, 40 )[c]{\huge $\times$}
  \Text(  36, 40 )[r]{$\mu_R$}
  \Text( 184, 40 )[l]{$\mu_L$}
  \Text(  80, 31 )[c]{$Y_\mu$}
  \Text( 140, 31 )[c]{$Y_\mu$}
  \Text( 110, 83 )[c]{$\Higgsino_d^0$}
  \Text( 111, 28 )[c]{$\smu_L$}
  \Text(   0, 40 )[c]{$a_\mu^{(i-R)}$:}
%
  \DashLine(   310, 40 )( 370, 40 ){4}
  \PhotonArc(  340, 40 )( 30, 0, 180 ){5}{8}
  \CArc(       340, 40 )( 30, 0, 180 )
  \Line( 270, 40 )( 310, 40 )
  \Line( 370, 40 )( 410, 40 )
  \Text( 390, 40 )[c]{\huge $\times$}
  \Text( 266, 40 )[r]{$\mu_R$}
  \Text( 414, 40 )[l]{$e_L$}
  \Text( 310, 31 )[c]{$Y_\mu$}
  \Text( 370, 31 )[c]{$Y_e$}
  \Text( 340, 83 )[c]{$\Higgsino_d^0$}
  \Text( 341.3, 40 )[c]{{\large $\otimes$}}
  \Text( 330, 28 )[c]{$\smu_L$}
  \Text( 352, 28 )[c]{$\se_L$}
  \Text( 230, 40 )[c]{$a_{\mu e \ph}^{(i-R)}$:}
\end{picture}
\begin{picture}(440,70)
%
  \DashLine(    80, 40 )( 140, 40 ){4}
  \PhotonArc(  110, 40 )( 30, 0, 180 ){5}{8}
  \CArc(       110, 40 )( 30, 0, 180 )
  \Line(  40, 40 )( 80, 40 )
  \Line( 140, 40 )( 180, 40 )
  \Text( 110, 70 )[c]{\huge $\times$}
  \Text(  36, 40 )[r]{$\mu_R$}
  \Text( 184, 40 )[l]{$\mu_L$}
  \Text(  80, 31 )[c]{$Y_\mu$}
  \Text( 140, 32 )[c]{$g_1$}
  \Text(  90, 80 )[c]{$\Higgsino_d^0$}
  \Text( 130, 80 )[c]{$\Bino$}
  \Text( 111, 28 )[c]{$\smu_L$}
  \Text(   0, 40 )[c]{$a_\mu^{(j-R)}$:}
%
  \DashLine(   310, 40 )( 370, 40 ){4}
  \PhotonArc(  340, 40 )( 30, 0, 180 ){5}{8}
  \CArc(       340, 40 )( 30, 0, 180 )
  \Line( 270, 40 )( 310, 40 )
  \Line( 370, 40 )( 410, 40 )
  \Text( 340, 70 )[c]{\huge $\times$}
  \Text( 266, 40 )[r]{$\mu_R$}
  \Text( 414, 40 )[l]{$e_L$}
  \Text( 310, 31 )[c]{$Y_\mu$}
  \Text( 374, 32 )[c]{$g_1$}
  \Text( 320, 80 )[c]{$\Higgsino_d^0$}
  \Text( 360, 80 )[c]{$\Bino$}
  \Text( 341.3, 40 )[c]{{\large $\otimes$}}
  \Text( 330, 28 )[c]{$\smu_L$}
  \Text( 354, 28 )[c]{$\se_L$}
  \Text( 230, 40 )[c]{$a_{\mu e \ph}^{(j-R)}$:}
\end{picture}
\begin{picture}(440,70)
%
  \DashLine(    80, 40 )( 140, 40 ){4}
  \PhotonArc(  110, 40 )( 30, 0, 180 ){5}{8}
  \CArc(       110, 40 )( 30, 0, 180 )
  \Line(  40, 40 )( 80, 40 )
  \Line( 140, 40 )( 180, 40 )
  \Text( 110, 70 )[c]{\huge $\times$}
  \Text(  36, 40 )[r]{$\mu_R$}
  \Text( 184, 40 )[l]{$\mu_L$}
  \Text(  80, 32 )[c]{$g_1$}
  \Text( 140, 31 )[c]{$Y_\mu$}
  \Text(  90, 80 )[c]{$\Bino$}
  \Text( 130, 80 )[c]{$\Higgsino_d^0$}
  \Text( 111, 28 )[c]{$\smu_R$}
  \Text(   0, 40 )[c]{$a_\mu^{(k-R)}$:}
%
  \DashLine(   310, 40 )( 370, 40 ){4}
  \PhotonArc(  340, 40 )( 30, 0, 180 ){5}{8}
  \CArc(       340, 40 )( 30, 0, 180 )
  \Line( 270, 40 )( 310, 40 )
  \Line( 370, 40 )( 410, 40 )
  \Text( 340, 70 )[c]{\huge $\times$}
  \Text( 266, 40 )[r]{$\mu_R$}
  \Text( 414, 40 )[l]{$e_L$}
  \Text( 310, 32 )[c]{$g_1$}
  \Text( 374, 31 )[c]{$Y_e$}
  \Text( 320, 80 )[c]{$\Bino$}
  \Text( 360, 80 )[c]{$\Higgsino_d^0$}
  \Text( 341.3, 40 )[c]{{\large $\otimes$}}
  \Text( 330, 28 )[c]{$\smu_R$}
  \Text( 354, 28 )[c]{$\se_R$}
  \Text( 230, 40 )[c]{$a_{\mu e \ph}^{(k-R)}$:}
\end{picture}
\begin{picture}(440,70)
%
  \DashLine(    80, 40 )( 140, 40 ){4}
  \PhotonArc(  110, 40 )( 30, 0, 180 ){5}{8}
  \CArc(       110, 40 )( 30, 0, 180 )
  \Line(  40, 40 )( 80, 40 )
  \Line( 140, 40 )( 180, 40 )
  \Text( 110, 70 )[c]{\huge $\times$}
  \Text(  36, 40 )[r]{$\mu_R$}
  \Text( 184, 40 )[l]{$\mu_L$}
  \Text(  80, 31 )[c]{$Y_\mu$}
  \Text( 140, 32 )[c]{$g_2$}
  \Text(  90, 80 )[c]{$\Higgsino_d^0$}
  \Text( 130, 80 )[c]{$\Wino^0$}
  \Text( 111, 28 )[c]{$\smu_L$}
  \Text(   0, 40 )[c]{$a_\mu^{(l-R)}$:}
%
  \DashLine(   310, 40 )( 370, 40 ){4}
  \PhotonArc(  340, 40 )( 30, 0, 180 ){5}{8}
  \CArc(       340, 40 )( 30, 0, 180 )
  \Line( 270, 40 )( 310, 40 )
  \Line( 370, 40 )( 410, 40 )
  \Text( 340, 70 )[c]{\huge $\times$}
  \Text( 266, 40 )[r]{$\mu_R$}
  \Text( 414, 40 )[l]{$e_L$}
  \Text( 310, 31 )[c]{$Y_\mu$}
  \Text( 370, 32 )[c]{$g_2$}
  \Text( 320, 80 )[c]{$\Higgsino_d^0$}
  \Text( 360, 80 )[c]{$\Wino^0$}
  \Text( 341.3, 40 )[c]{{\large $\otimes$}}
  \Text( 330, 28 )[c]{$\smu_L$}
  \Text( 354, 28 )[c]{$\se_L$}
  \Text( 230, 40 )[c]{$a_{\mu e \ph}^{(l-R)}$:}
\end{picture}
\vspace*{-0.8cm}
\caption{Neutralino-slepton contributions to $a_r$ 
that give rise to muon $g-2$ and $\mu \ra e \ph$ in the interaction 
eigenstate basis.  
The photon (not shown) is emitted from the slepton.
The chirality flip is shown by a $\times$
on the fermion line and the slepton flavor violating mass insertion
is shown by a $\otimes$.}
\label{neut-R-fig}
\end{figure}

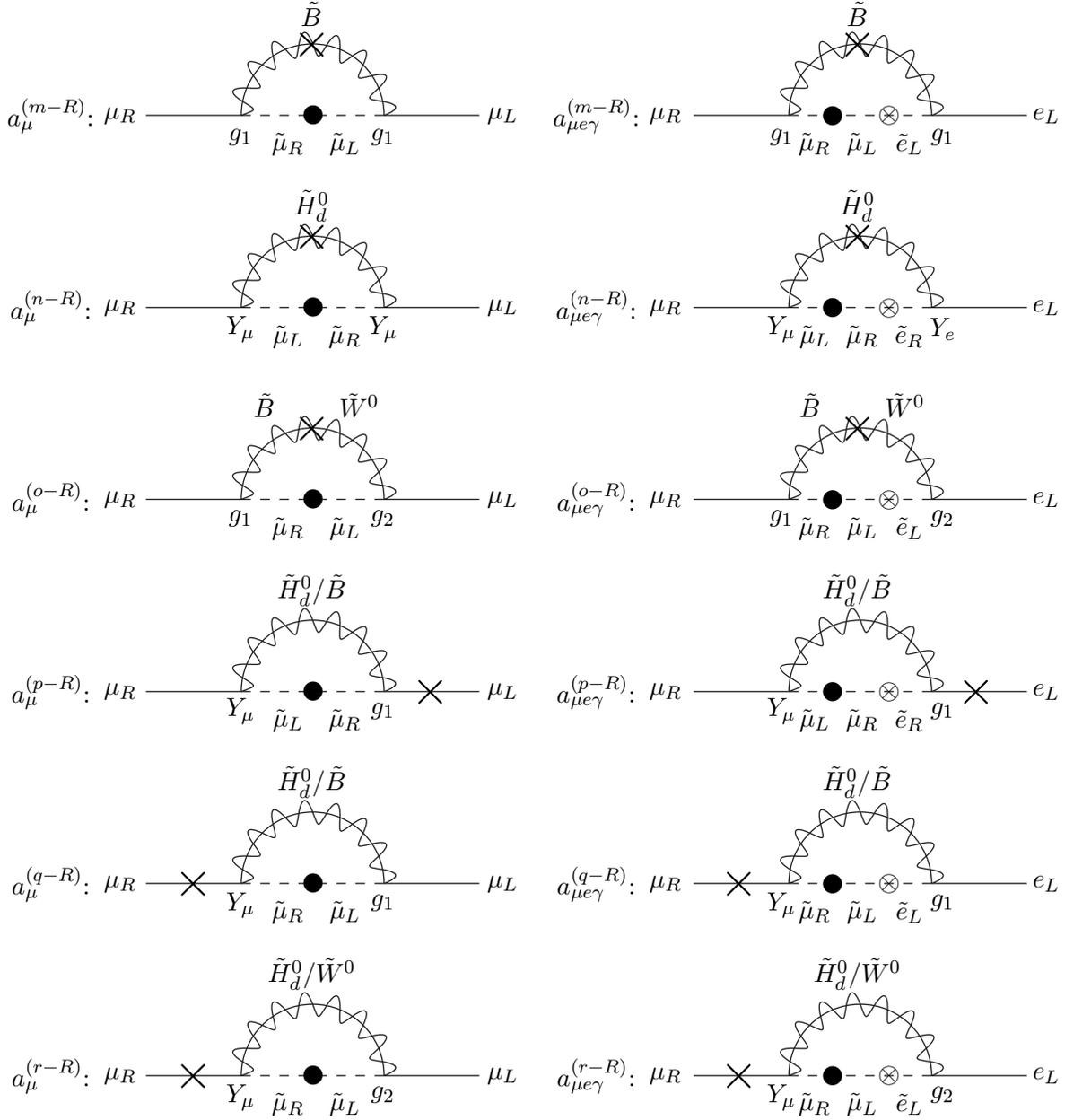
\begin{figure}[t]
\begin{picture}(440,80)
%
  \DashLine(    80, 40 )( 140, 40 ){4}
  \PhotonArc(  110, 40 )( 30, 0, 180 ){5}{8}
  \CArc(       110, 40 )( 30, 0, 180 )
  \Line(  40, 40 )( 80, 40 )
  \Line( 140, 40 )( 180, 40 )
  \Text( 110, 70 )[c]{\huge $\times$}
  \Text( 110.6, 39.8 )[c]{\huge $\bullet$}
  \Text(  36, 40 )[r]{$\mu_R$}
  \Text( 184, 40 )[l]{$\mu_L$}
  \Text(  80, 31 )[c]{$g_1$}
  \Text( 140, 31 )[c]{$g_1$}
  \Text( 110, 83 )[c]{$\Bino$}
  \Text( 100, 28 )[c]{$\smu_R$}
  \Text( 124, 28 )[c]{$\smu_L$}
  \Text(   0, 40 )[c]{$a_\mu^{(m-R)}$:}
%
  \DashLine(   310, 40 )( 370, 40 ){4}
  \PhotonArc(  340, 40 )( 30, 0, 180 ){5}{8}
  \CArc(       340, 40 )( 30, 0, 180 )
  \Line( 270, 40 )( 310, 40 )
  \Line( 370, 40 )( 410, 40 )
  \Text( 340, 70 )[c]{\huge $\times$}
  \Text( 329.5, 39.5 )[c]{\huge $\bullet$}
  \Text( 353.5, 40 )[c]{\large $\otimes$}
  \Text( 266, 40 )[r]{$\mu_R$}
  \Text( 414, 40 )[l]{$e_L$}
  \Text( 308, 31 )[c]{$g_1$}
  \Text( 376, 31 )[c]{$g_1$}
  \Text( 340, 83 )[c]{$\Bino$}
  \Text( 322, 28 )[c]{$\smu_R$}
  \Text( 342, 28 )[c]{$\smu_L$}
  \Text( 362, 28 )[c]{$\se_L$}
  \Text( 230, 40 )[c]{$a_{\mu e \ph}^{(m-R)}$:}
\end{picture}
\begin{picture}(440,80)
%
  \DashLine(    80, 40 )( 140, 40 ){4}
  \PhotonArc(  110, 40 )( 30, 0, 180 ){5}{8}
  \CArc(       110, 40 )( 30, 0, 180 )
  \Line(  40, 40 )( 80, 40 )
  \Line( 140, 40 )( 180, 40 )
  \Text( 110, 70 )[c]{\huge $\times$}
  \Text( 110.6, 39.8 )[c]{\huge $\bullet$}
  \Text(  36, 40 )[r]{$\mu_R$}
  \Text( 184, 40 )[l]{$\mu_L$}
  \Text(  80, 31 )[c]{$Y_\mu$}
  \Text( 140, 31 )[c]{$Y_\mu$}
  \Text( 110, 83 )[c]{$\Higgsino_d^0$}
  \Text( 100, 28 )[c]{$\smu_L$}
  \Text( 124, 28 )[c]{$\smu_R$}
  \Text(   0, 40 )[c]{$a_\mu^{(n-R)}$:}
%
  \DashLine(   310, 40 )( 370, 40 ){4}
  \PhotonArc(  340, 40 )( 30, 0, 180 ){5}{8}
  \CArc(       340, 40 )( 30, 0, 180 )
  \Line( 270, 40 )( 310, 40 )
  \Line( 370, 40 )( 410, 40 )
  \Text( 340, 70 )[c]{\huge $\times$}
  \Text( 329.5, 39.5 )[c]{\huge $\bullet$}
  \Text( 353.5, 40 )[c]{\large $\otimes$}
  \Text( 266, 40 )[r]{$\mu_R$}
  \Text( 414, 40 )[l]{$e_L$}
  \Text( 308, 31 )[c]{$Y_\mu$}
  \Text( 376, 31 )[c]{$Y_e$}
  \Text( 340, 83 )[c]{$\Higgsino_d^0$}
  \Text( 322, 28 )[c]{$\smu_L$}
  \Text( 342, 28 )[c]{$\smu_R$}
  \Text( 362, 28 )[c]{$\se_R$}
  \Text( 230, 40 )[c]{$a_{\mu e \ph}^{(n-R)}$:}
\end{picture}
\begin{picture}(440,80)
%
  \DashLine(    80, 40 )( 140, 40 ){4}
  \PhotonArc(  110, 40 )( 30, 0, 180 ){5}{8}
  \CArc(       110, 40 )( 30, 0, 180 )
  \Line(  40, 40 )( 80, 40 )
  \Line( 140, 40 )( 180, 40 )
  \Text( 110, 70 )[c]{\huge $\times$}
  \Text( 110.6, 39.8 )[c]{\huge $\bullet$}
  \Text(  36, 40 )[r]{$\mu_R$}
  \Text( 184, 40 )[l]{$\mu_L$}
  \Text(  80, 32 )[c]{$g_1$}
  \Text( 140, 32 )[c]{$g_2$}
  \Text(  90, 80 )[c]{$\Bino$}
  \Text( 130, 80 )[c]{$\Wino^0$}
  \Text( 100, 28 )[c]{$\smu_R$}
  \Text( 124, 28 )[c]{$\smu_L$}
  \Text(   0, 40 )[c]{$a_\mu^{(o-R)}$:}
%
  \DashLine(   310, 40 )( 370, 40 ){4}
  \PhotonArc(  340, 40 )( 30, 0, 180 ){5}{8}
  \CArc(       340, 40 )( 30, 0, 180 )
  \Line( 270, 40 )( 310, 40 )
  \Line( 370, 40 )( 410, 40 )
  \Text( 340, 70 )[c]{\huge $\times$}
  \Text( 329.5, 39.5 )[c]{\huge $\bullet$}
  \Text( 353.5, 40 )[c]{\large $\otimes$}
  \Text( 266, 40 )[r]{$\mu_R$}
  \Text( 414, 40 )[l]{$e_L$}
  \Text( 308, 32 )[c]{$g_1$}
  \Text( 376, 32 )[c]{$g_2$}
  \Text( 320, 80 )[c]{$\Bino$}
  \Text( 360, 80 )[c]{$\Wino^0$}
  \Text( 322, 28 )[c]{$\smu_R$}
  \Text( 342, 28 )[c]{$\smu_L$}
  \Text( 362, 28 )[c]{$\se_L$}
  \Text( 230, 40 )[c]{$a_{\mu e \ph}^{(o-R)}$:}
\end{picture}
\begin{picture}(440,80)
%
  \DashLine(    80, 40 )( 140, 40 ){4}
  \PhotonArc(  110, 40 )( 30, 0, 180 ){5}{8}
  \CArc(       110, 40 )( 30, 0, 180 )
  \Line(  40, 40 )( 80, 40 )
  \Line( 140, 40 )( 180, 40 )
  \Text( 160, 40 )[c]{\huge $\times$}
  \Text( 110.6, 39.8 )[c]{\huge $\bullet$}
  \Text(  36, 40 )[r]{$\mu_R$}
  \Text( 184, 40 )[l]{$\mu_L$}
  \Text(  80, 32 )[c]{$Y_\mu$}
  \Text( 140, 32 )[c]{$g_1$}
  \Text( 110, 83 )[c]{$\Higgsino_d^0/\Bino$}
  \Text( 100, 28 )[c]{$\smu_L$}
  \Text( 124, 28 )[c]{$\smu_R$}
  \Text(   0, 40 )[c]{$a_\mu^{(p-R)}$:}
%
  \DashLine(   310, 40 )( 370, 40 ){4}
  \PhotonArc(  340, 40 )( 30, 0, 180 ){5}{8}
  \CArc(       340, 40 )( 30, 0, 180 )
  \Line( 270, 40 )( 310, 40 )
  \Line( 370, 40 )( 410, 40 )
  \Text( 390, 40 )[c]{\huge $\times$}
  \Text( 329.5, 39.5 )[c]{\huge $\bullet$}
  \Text( 353.5, 40 )[c]{\large $\otimes$}
  \Text( 266, 40 )[r]{$\mu_R$}
  \Text( 414, 40 )[l]{$e_L$}
  \Text( 308, 32 )[c]{$Y_\mu$}
  \Text( 376, 32 )[c]{$g_1$}
  \Text( 340, 83 )[c]{$\Higgsino_d^0/\Bino$}
  \Text( 322, 28 )[c]{$\smu_L$}
  \Text( 342, 28 )[c]{$\smu_R$}
  \Text( 362, 28 )[c]{$\se_R$}
  \Text( 230, 40 )[c]{$a_{\mu e \ph}^{(p-R)}$:}
\end{picture}
\begin{picture}(440,80)
%
  \DashLine(    80, 40 )( 140, 40 ){4}
  \PhotonArc(  110, 40 )( 30, 0, 180 ){5}{8}
  \CArc(       110, 40 )( 30, 0, 180 )
  \Line(  40, 40 )( 80, 40 )
  \Line( 140, 40 )( 180, 40 )
  \Text(  60, 40 )[c]{\huge $\times$}
  \Text( 110.6, 39.8 )[c]{\huge $\bullet$}
  \Text(  36, 40 )[r]{$\mu_R$}
  \Text( 184, 40 )[l]{$\mu_L$}
  \Text(  80, 31 )[c]{$Y_\mu$}
  \Text( 140, 32 )[c]{$g_1$}
  \Text( 110, 83 )[c]{$\Higgsino_d^0/\Bino$}
  \Text( 100, 28 )[c]{$\smu_R$}
  \Text( 124, 28 )[c]{$\smu_L$}
  \Text(   0, 40 )[c]{$a_\mu^{(q-R)}$:}
%
  \DashLine(   310, 40 )( 370, 40 ){4}
  \PhotonArc(  340, 40 )( 30, 0, 180 ){5}{8}
  \CArc(       340, 40 )( 30, 0, 180 )
  \Line( 270, 40 )( 310, 40 )
  \Line( 370, 40 )( 410, 40 )
  \Text( 290, 40 )[c]{\huge $\times$}
  \Text( 329.5, 39.5 )[c]{\huge $\bullet$}
  \Text( 353.5, 40 )[c]{\large $\otimes$}
  \Text( 266, 40 )[r]{$\mu_R$}
  \Text( 414, 40 )[l]{$e_L$}
  \Text( 308, 31 )[c]{$Y_\mu$}
  \Text( 376, 32 )[c]{$g_1$}
  \Text( 340, 83 )[c]{$\Higgsino_d^0/\Bino$}
  \Text( 322, 28 )[c]{$\smu_R$}
  \Text( 342, 28 )[c]{$\smu_L$}
  \Text( 362, 28 )[c]{$\se_L$}
  \Text( 230, 40 )[c]{$a_{\mu e \ph}^{(q-R)}$:}
\end{picture}
\begin{picture}(440,80)
%
  \DashLine(    80, 40 )( 140, 40 ){4}
  \PhotonArc(  110, 40 )( 30, 0, 180 ){5}{8}
  \CArc(       110, 40 )( 30, 0, 180 )
  \Line(  40, 40 )( 80, 40 )
  \Line( 140, 40 )( 180, 40 )
  \Text(  60, 40 )[c]{\huge $\times$}
  \Text( 110.6, 39.8 )[c]{\huge $\bullet$}
  \Text(  36, 40 )[r]{$\mu_R$}
  \Text( 184, 40 )[l]{$\mu_L$}
  \Text(  80, 31 )[c]{$Y_\mu$}
  \Text( 140, 32 )[c]{$g_2$}
  \Text( 110, 83 )[c]{$\Higgsino_d^0/\Wino^0$}
  \Text( 100, 28 )[c]{$\smu_R$}
  \Text( 124, 28 )[c]{$\smu_L$}
  \Text(   0, 40 )[c]{$a_\mu^{(r-R)}$:}
%
  \DashLine(   310, 40 )( 370, 40 ){4}
  \PhotonArc(  340, 40 )( 30, 0, 180 ){5}{8}
  \CArc(       340, 40 )( 30, 0, 180 )
  \Line( 270, 40 )( 310, 40 )
  \Line( 370, 40 )( 410, 40 )
  \Text( 290, 40 )[c]{\huge $\times$}
  \Text( 329.5, 39.5 )[c]{\huge $\bullet$}
  \Text( 353.5, 40 )[c]{\large $\otimes$}
  \Text( 266, 40 )[r]{$\mu_R$}
  \Text( 414, 40 )[l]{$e_L$}
  \Text( 308, 31 )[c]{$Y_\mu$}
  \Text( 376, 32 )[c]{$g_2$}
  \Text( 340, 83 )[c]{$\Higgsino_d^0/\Wino^0$}
  \Text( 322, 28 )[c]{$\smu_R$}
  \Text( 342, 28 )[c]{$\smu_L$}
  \Text( 362, 28 )[c]{$\se_L$}
  \Text( 230, 40 )[c]{$a_{\mu e \ph}^{(r-R)}$:}
\end{picture}
\caption{Similar to Fig.~\ref{neut-R-fig}, the neutralino-slepton 
contributions to $a_r$
with a left-right slepton mass insertion
that give rise to muon $g-2$ and $\mu \ra e \ph$ in the interaction 
eigenstate basis are shown.  The slepton left-right mass insertion
is shown by a $\bullet$.}
\label{neut-R-LR-fig}
\end{figure}

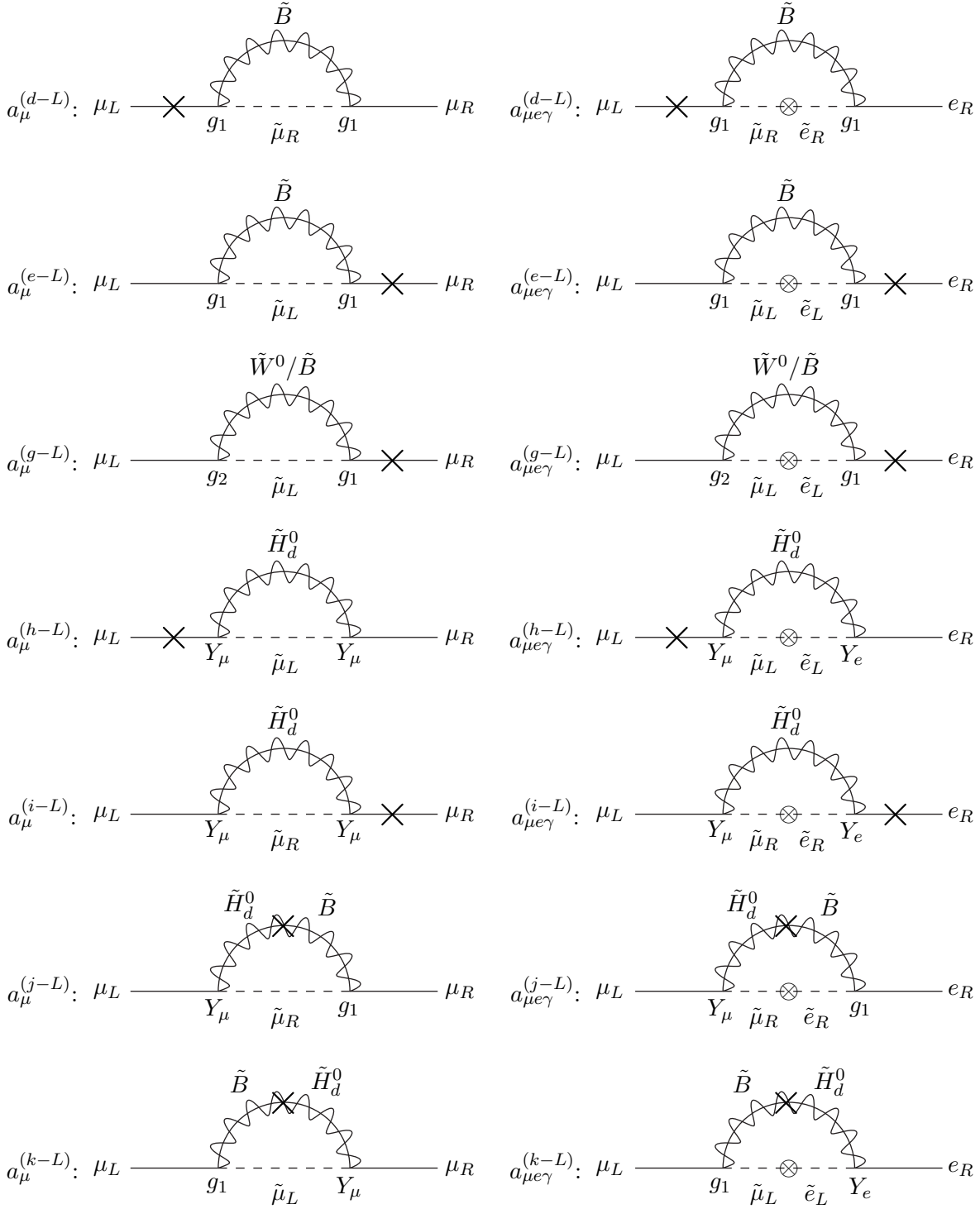
\begin{figure}[t]
\begin{picture}(440,80)
%
  \DashLine(    80, 40 )( 140, 40 ){4}
  \PhotonArc(  110, 40 )( 30, 0, 180 ){5}{8}
  \CArc(       110, 40 )( 30, 0, 180 )
  \Line(  40, 40 )( 80, 40 )
  \Line( 140, 40 )( 180, 40 )
  \Text(  60, 40 )[c]{\huge $\times$}
  \Text(  36, 40 )[r]{$\mu_L$}
  \Text( 184, 40 )[l]{$\mu_R$}
  \Text(  80, 32 )[c]{$g_1$}
  \Text( 140, 32 )[c]{$g_1$}
  \Text( 110, 83 )[c]{$\Bino$}
  \Text( 111, 28 )[c]{$\smu_R$}
  \Text(   0, 40 )[c]{$a_\mu^{(d-L)}$:}
%
  \DashLine(   310, 40 )( 370, 40 ){4}
  \PhotonArc(  340, 40 )( 30, 0, 180 ){5}{8}
  \CArc(       340, 40 )( 30, 0, 180 )
  \Line( 270, 40 )( 310, 40 )
  \Line( 370, 40 )( 410, 40 )
  \Text( 290, 40 )[c]{\huge $\times$}
  \Text( 266, 40 )[r]{$\mu_L$}
  \Text( 414, 40 )[l]{$e_R$}
  \Text( 310, 32 )[c]{$g_1$}
  \Text( 370, 32 )[c]{$g_1$}
  \Text( 340, 83 )[c]{$\Bino$}
  \Text( 341.3, 40 )[c]{{\large $\otimes$}}
  \Text( 330, 28 )[c]{$\smu_R$}
  \Text( 352, 28 )[c]{$\se_R$}
  \Text( 230, 40 )[c]{$a_{\mu e \ph}^{(d-L)}$:}
\end{picture}
\begin{picture}(440,80)
%
  \DashLine(    80, 40 )( 140, 40 ){4}
  \PhotonArc(  110, 40 )( 30, 0, 180 ){5}{8}
  \CArc(       110, 40 )( 30, 0, 180 )
  \Line(  40, 40 )( 80, 40 )
  \Line( 140, 40 )( 180, 40 )
  \Text( 160, 40 )[c]{\huge $\times$}
  \Text(  36, 40 )[r]{$\mu_L$}
  \Text( 184, 40 )[l]{$\mu_R$}
  \Text(  80, 32 )[c]{$g_1$}
  \Text( 140, 32 )[c]{$g_1$}
  \Text( 110, 83 )[c]{$\Bino$}
  \Text( 111, 28 )[c]{$\smu_L$}
  \Text(   0, 40 )[c]{$a_\mu^{(e-L)}$:}
%
  \DashLine(   310, 40 )( 370, 40 ){4}
  \PhotonArc(  340, 40 )( 30, 0, 180 ){5}{8}
  \CArc(       340, 40 )( 30, 0, 180 )
  \Line( 270, 40 )( 310, 40 )
  \Line( 370, 40 )( 410, 40 )
  \Text( 390, 40 )[c]{\huge $\times$}
  \Text( 266, 40 )[r]{$\mu_L$}
  \Text( 414, 40 )[l]{$e_R$}
  \Text( 310, 32 )[c]{$g_1$}
  \Text( 370, 32 )[c]{$g_1$}
  \Text( 340, 83 )[c]{$\Bino$}
  \Text( 341.3, 40 )[c]{{\large $\otimes$}}
  \Text( 330, 28 )[c]{$\smu_L$}
  \Text( 352, 28 )[c]{$\se_L$}
  \Text( 230, 40 )[c]{$a_{\mu e \ph}^{(e-L)}$:}
\end{picture}
\begin{picture}(440,80)
%
  \DashLine(    80, 40 )( 140, 40 ){4}
  \PhotonArc(  110, 40 )( 30, 0, 180 ){5}{8}
  \CArc(       110, 40 )( 30, 0, 180 )
  \Line(  40, 40 )( 80, 40 )
  \Line( 140, 40 )( 180, 40 )
  \Text( 160, 40 )[c]{\huge $\times$}
  \Text(  36, 40 )[r]{$\mu_L$}
  \Text( 184, 40 )[l]{$\mu_R$}
  \Text(  80, 32 )[c]{$g_2$}
  \Text( 140, 32 )[c]{$g_1$}
  \Text( 110, 83 )[c]{$\Wino^0/\Bino$}
  \Text( 111, 28 )[c]{$\smu_L$}
  \Text(   0, 40 )[c]{$a_\mu^{(g-L)}$:}
%
  \DashLine(   310, 40 )( 370, 40 ){4}
  \PhotonArc(  340, 40 )( 30, 0, 180 ){5}{8}
  \CArc(       340, 40 )( 30, 0, 180 )
  \Line( 270, 40 )( 310, 40 )
  \Line( 370, 40 )( 410, 40 )
  \Text( 390, 40 )[c]{\huge $\times$}
  \Text( 266, 40 )[r]{$\mu_L$}
  \Text( 414, 40 )[l]{$e_R$}
  \Text( 310, 32 )[c]{$g_2$}
  \Text( 370, 32 )[c]{$g_1$}
  \Text( 340, 83 )[c]{$\Wino^0/\Bino$}
  \Text( 341.3, 40 )[c]{{\large $\otimes$}}
  \Text( 330, 28 )[c]{$\smu_L$}
  \Text( 352, 28 )[c]{$\se_L$}
  \Text( 230, 40 )[c]{$a_{\mu e \ph}^{(g-L)}$:}
\end{picture}
\begin{picture}(440,80)
%
  \DashLine(    80, 40 )( 140, 40 ){4}
  \PhotonArc(  110, 40 )( 30, 0, 180 ){5}{8}
  \CArc(       110, 40 )( 30, 0, 180 )
  \Line(  40, 40 )( 80, 40 )
  \Line( 140, 40 )( 180, 40 )
  \Text(  60, 40 )[c]{\huge $\times$}
  \Text(  36, 40 )[r]{$\mu_L$}
  \Text( 184, 40 )[l]{$\mu_R$}
  \Text(  80, 32 )[c]{$Y_\mu$}
  \Text( 140, 32 )[c]{$Y_\mu$}
  \Text( 110, 83 )[c]{$\Higgsino_d^0$}
  \Text( 111, 28 )[c]{$\smu_L$}
  \Text(   0, 40 )[c]{$a_\mu^{(h-L)}$:}
%
  \DashLine(   310, 40 )( 370, 40 ){4}
  \PhotonArc(  340, 40 )( 30, 0, 180 ){5}{8}
  \CArc(       340, 40 )( 30, 0, 180 )
  \Line( 270, 40 )( 310, 40 )
  \Line( 370, 40 )( 410, 40 )
  \Text( 290, 40 )[c]{\huge $\times$}
  \Text( 266, 40 )[r]{$\mu_L$}
  \Text( 414, 40 )[l]{$e_R$}
  \Text( 310, 32 )[c]{$Y_\mu$}
  \Text( 370, 32 )[c]{$Y_e$}
  \Text( 340, 83 )[c]{$\Higgsino_d^0$}
  \Text( 341.3, 40 )[c]{{\large $\otimes$}}
  \Text( 330, 28 )[c]{$\smu_L$}
  \Text( 352, 28 )[c]{$\se_L$}
  \Text( 230, 40 )[c]{$a_{\mu e \ph}^{(h-L)}$:}
\end{picture}
\begin{picture}(440,80)
%
  \DashLine(    80, 40 )( 140, 40 ){4}
  \PhotonArc(  110, 40 )( 30, 0, 180 ){5}{8}
  \CArc(       110, 40 )( 30, 0, 180 )
  \Line(  40, 40 )( 80, 40 )
  \Line( 140, 40 )( 180, 40 )
  \Text( 160, 40 )[c]{\huge $\times$}
  \Text(  36, 40 )[r]{$\mu_L$}
  \Text( 184, 40 )[l]{$\mu_R$}
  \Text(  80, 32 )[c]{$Y_\mu$}
  \Text( 140, 32 )[c]{$Y_\mu$}
  \Text( 110, 83 )[c]{$\Higgsino_d^0$}
  \Text( 111, 28 )[c]{$\smu_R$}
  \Text(   0, 40 )[c]{$a_\mu^{(i-L)}$:}
%
  \DashLine(   310, 40 )( 370, 40 ){4}
  \PhotonArc(  340, 40 )( 30, 0, 180 ){5}{8}
  \CArc(       340, 40 )( 30, 0, 180 )
  \Line( 270, 40 )( 310, 40 )
  \Line( 370, 40 )( 410, 40 )
  \Text( 390, 40 )[c]{\huge $\times$}
  \Text( 266, 40 )[r]{$\mu_L$}
  \Text( 414, 40 )[l]{$e_R$}
  \Text( 310, 32 )[c]{$Y_\mu$}
  \Text( 370, 32 )[c]{$Y_e$}
  \Text( 340, 83 )[c]{$\Higgsino_d^0$}
  \Text( 341.3, 40 )[c]{{\large $\otimes$}}
  \Text( 330, 28 )[c]{$\smu_R$}
  \Text( 352, 28 )[c]{$\se_R$}
  \Text( 230, 40 )[c]{$a_{\mu e \ph}^{(i-L)}$:}
\end{picture}
\begin{picture}(440,80)
%
  \DashLine(    80, 40 )( 140, 40 ){4}
  \PhotonArc(  110, 40 )( 30, 0, 180 ){5}{8}
  \CArc(       110, 40 )( 30, 0, 180 )
  \Line(  40, 40 )( 80, 40 )
  \Line( 140, 40 )( 180, 40 )
  \Text( 110, 70 )[c]{\huge $\times$}
  \Text(  36, 40 )[r]{$\mu_L$}
  \Text( 184, 40 )[l]{$\mu_R$}
  \Text(  80, 31 )[c]{$Y_\mu$}
  \Text( 140, 32 )[c]{$g_1$}
  \Text(  90, 80 )[c]{$\Higgsino_d^0$}
  \Text( 130, 80 )[c]{$\Bino$}
  \Text( 111, 28 )[c]{$\smu_R$}
  \Text(   0, 40 )[c]{$a_\mu^{(j-L)}$:}
%
  \DashLine(   310, 40 )( 370, 40 ){4}
  \PhotonArc(  340, 40 )( 30, 0, 180 ){5}{8}
  \CArc(       340, 40 )( 30, 0, 180 )
  \Line( 270, 40 )( 310, 40 )
  \Line( 370, 40 )( 410, 40 )
  \Text( 340, 70 )[c]{\huge $\times$}
  \Text( 266, 40 )[r]{$\mu_L$}
  \Text( 414, 40 )[l]{$e_R$}
  \Text( 310, 31 )[c]{$Y_\mu$}
  \Text( 374, 32 )[c]{$g_1$}
  \Text( 320, 80 )[c]{$\Higgsino_d^0$}
  \Text( 360, 80 )[c]{$\Bino$}
  \Text( 341.3, 40 )[c]{{\large $\otimes$}}
  \Text( 330, 28 )[c]{$\smu_R$}
  \Text( 354, 28 )[c]{$\se_R$}
  \Text( 230, 40 )[c]{$a_{\mu e \ph}^{(j-L)}$:}
\end{picture}
\begin{picture}(440,80)
%
  \DashLine(    80, 40 )( 140, 40 ){4}
  \PhotonArc(  110, 40 )( 30, 0, 180 ){5}{8}
  \CArc(       110, 40 )( 30, 0, 180 )
  \Line(  40, 40 )( 80, 40 )
  \Line( 140, 40 )( 180, 40 )
  \Text( 110, 70 )[c]{\huge $\times$}
  \Text(  36, 40 )[r]{$\mu_L$}
  \Text( 184, 40 )[l]{$\mu_R$}
  \Text(  80, 32 )[c]{$g_1$}
  \Text( 140, 31 )[c]{$Y_\mu$}
  \Text(  90, 80 )[c]{$\Bino$}
  \Text( 130, 80 )[c]{$\Higgsino_d^0$}
  \Text( 111, 28 )[c]{$\smu_L$}
  \Text(   0, 40 )[c]{$a_\mu^{(k-L)}$:}
%
  \DashLine(   310, 40 )( 370, 40 ){4}
  \PhotonArc(  340, 40 )( 30, 0, 180 ){5}{8}
  \CArc(       340, 40 )( 30, 0, 180 )
  \Line( 270, 40 )( 310, 40 )
  \Line( 370, 40 )( 410, 40 )
  \Text( 340, 70 )[c]{\huge $\times$}
  \Text( 266, 40 )[r]{$\mu_L$}
  \Text( 414, 40 )[l]{$e_R$}
  \Text( 310, 32 )[c]{$g_1$}
  \Text( 374, 31 )[c]{$Y_e$}
  \Text( 320, 80 )[c]{$\Bino$}
  \Text( 360, 80 )[c]{$\Higgsino_d^0$}
  \Text( 341.3, 40 )[c]{{\large $\otimes$}}
  \Text( 330, 28 )[c]{$\smu_L$}
  \Text( 354, 28 )[c]{$\se_L$}
  \Text( 230, 40 )[c]{$a_{\mu e \ph}^{(k-L)}$:}
\end{picture}
\caption{Neutralino-slepton contributions to $a_l$ 
that give rise to muon $g-2$ and $\mu \ra e \ph$ in the interaction 
eigenstate basis.  The chirality flip is shown by a $\times$
on the fermion line and the slepton flavor violating mass insertion
is shown by a $\otimes$.}
\label{neut-L-fig}
\end{figure}

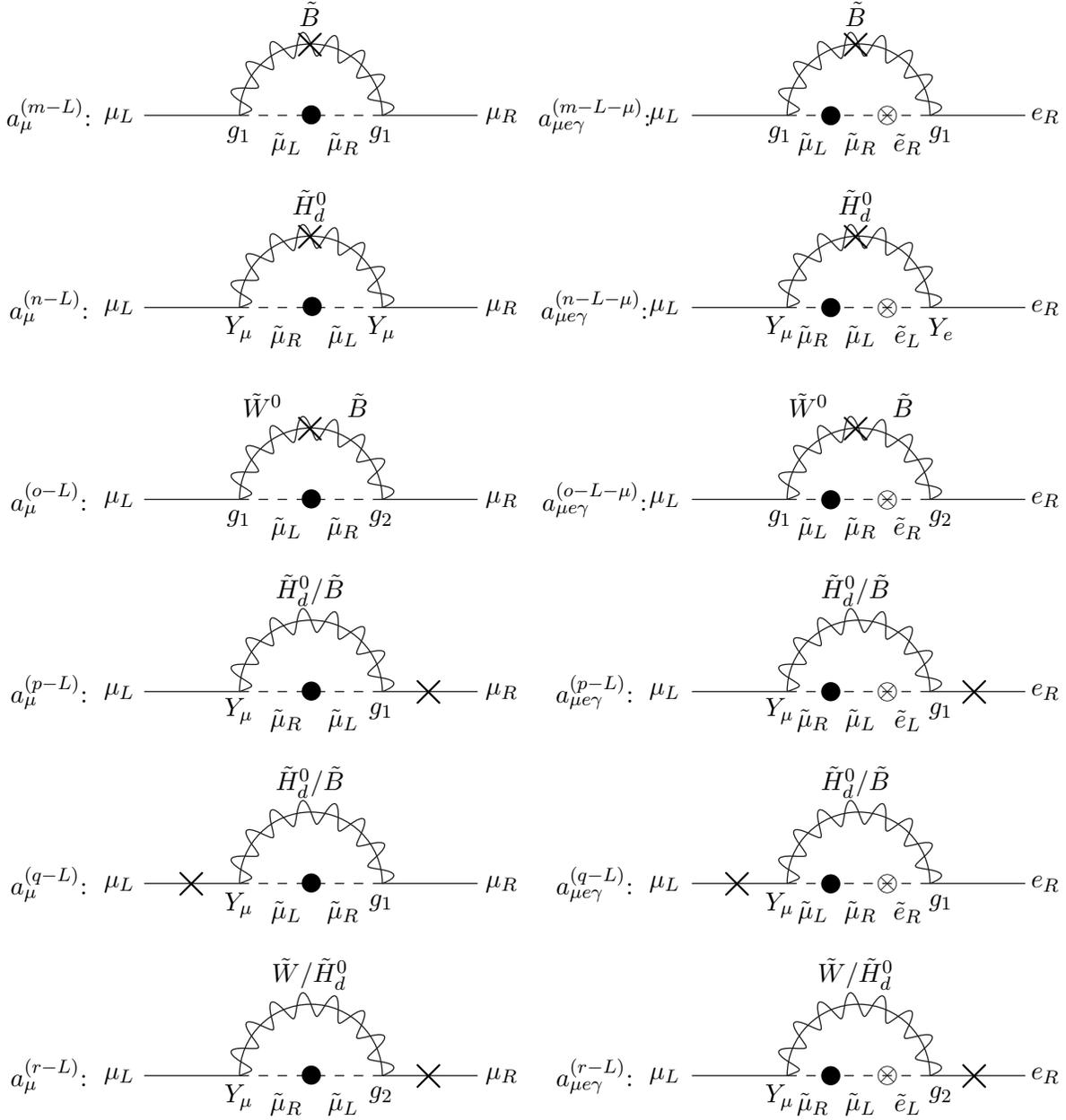
\begin{figure}[t]
\begin{picture}(440,80)
%
  \DashLine(    80, 40 )( 140, 40 ){4}
  \PhotonArc(  110, 40 )( 30, 0, 180 ){5}{8}
  \CArc(       110, 40 )( 30, 0, 180 )
  \Line(  40, 40 )( 80, 40 )
  \Line( 140, 40 )( 180, 40 )
  \Text( 110, 70 )[c]{\huge $\times$}
  \Text( 110.6, 39.8 )[c]{\huge $\bullet$}
  \Text(  36, 40 )[r]{$\mu_L$}
  \Text( 184, 40 )[l]{$\mu_R$}
  \Text(  80, 32 )[c]{$g_1$}
  \Text( 140, 32 )[c]{$g_1$}
  \Text( 110, 83 )[c]{$\Bino$}
  \Text( 100, 28 )[c]{$\smu_L$}
  \Text( 124, 28 )[c]{$\smu_R$}
  \Text(   0, 40 )[c]{$a_\mu^{(m-L)}$:}
%
  \DashLine(   310, 40 )( 370, 40 ){4}
  \PhotonArc(  340, 40 )( 30, 0, 180 ){5}{8}
  \CArc(       340, 40 )( 30, 0, 180 )
  \Line( 270, 40 )( 310, 40 )
  \Line( 370, 40 )( 410, 40 )
  \Text( 340, 70 )[c]{\huge $\times$}
  \Text( 329.5, 39.5 )[c]{\huge $\bullet$}
  \Text( 353.5, 40 )[c]{\large $\otimes$}
  \Text( 266, 40 )[r]{$\mu_L$}
  \Text( 414, 40 )[l]{$e_R$}
  \Text( 308, 32 )[c]{$g_1$}
  \Text( 376, 32 )[c]{$g_1$}
  \Text( 340, 83 )[c]{$\Bino$}
  \Text( 322, 28 )[c]{$\smu_L$}
  \Text( 342, 28 )[c]{$\smu_R$}
  \Text( 362, 28 )[c]{$\se_R$}
  \Text( 230, 40 )[c]{$a_{\mu e \ph}^{(m-L-\mu)}$:}
\end{picture}
\begin{picture}(440,80)
%
  \DashLine(    80, 40 )( 140, 40 ){4}
  \PhotonArc(  110, 40 )( 30, 0, 180 ){5}{8}
  \CArc(       110, 40 )( 30, 0, 180 )
  \Line(  40, 40 )( 80, 40 )
  \Line( 140, 40 )( 180, 40 )
  \Text( 110, 70 )[c]{\huge $\times$}
  \Text( 111, 40 )[c]{\huge $\bullet$}
  \Text(  36, 40 )[r]{$\mu_L$}
  \Text( 184, 40 )[l]{$\mu_R$}
  \Text(  80, 31 )[c]{$Y_\mu$}
  \Text( 140, 31 )[c]{$Y_\mu$}
  \Text( 110, 83 )[c]{$\Higgsino_d^0$}
  \Text( 100, 28 )[c]{$\smu_R$}
  \Text( 124, 28 )[c]{$\smu_L$}
  \Text(   0, 40 )[c]{$a_\mu^{(n-L)}$:}
%
  \DashLine(   310, 40 )( 370, 40 ){4}
  \PhotonArc(  340, 40 )( 30, 0, 180 ){5}{8}
  \CArc(       340, 40 )( 30, 0, 180 )
  \Line( 270, 40 )( 310, 40 )
  \Line( 370, 40 )( 410, 40 )
  \Text( 340, 70 )[c]{\huge $\times$}
  \Text( 329.5, 39.5 )[c]{\huge $\bullet$}
  \Text( 353.5, 40 )[c]{\large $\otimes$}
  \Text( 266, 40 )[r]{$\mu_L$}
  \Text( 414, 40 )[l]{$e_R$}
  \Text( 308, 31 )[c]{$Y_\mu$}
  \Text( 376, 31 )[c]{$Y_e$}
  \Text( 340, 83 )[c]{$\Higgsino_d^0$}
  \Text( 322, 28 )[c]{$\smu_R$}
  \Text( 342, 28 )[c]{$\smu_L$}
  \Text( 362, 28 )[c]{$\se_L$}
  \Text( 230, 40 )[c]{$a_{\mu e \ph}^{(n-L-\mu)}$:}
\end{picture}
\begin{picture}(440,80)
%
  \DashLine(    80, 40 )( 140, 40 ){4}
  \PhotonArc(  110, 40 )( 30, 0, 180 ){5}{8}
  \CArc(       110, 40 )( 30, 0, 180 )
  \Line(  40, 40 )( 80, 40 )
  \Line( 140, 40 )( 180, 40 )
  \Text( 110, 70 )[c]{\huge $\times$}
  \Text( 110.6, 39.8 )[c]{\huge $\bullet$}
  \Text(  36, 40 )[r]{$\mu_L$}
  \Text( 184, 40 )[l]{$\mu_R$}
  \Text(  80, 32 )[c]{$g_1$}
  \Text( 140, 32 )[c]{$g_2$}
  \Text(  90, 80 )[c]{$\Wino^0$}
  \Text( 130, 80 )[c]{$\Bino$}
  \Text( 100, 28 )[c]{$\smu_L$}
  \Text( 124, 28 )[c]{$\smu_R$}
  \Text(   0, 40 )[c]{$a_\mu^{(o-L)}$:}
%
  \DashLine(   310, 40 )( 370, 40 ){4}
  \PhotonArc(  340, 40 )( 30, 0, 180 ){5}{8}
  \CArc(       340, 40 )( 30, 0, 180 )
  \Line( 270, 40 )( 310, 40 )
  \Line( 370, 40 )( 410, 40 )
  \Text( 340, 70 )[c]{\huge $\times$}
  \Text( 329.5, 39.5 )[c]{\huge $\bullet$}
  \Text( 353.5, 40 )[c]{\large $\otimes$}
  \Text( 266, 40 )[r]{$\mu_L$}
  \Text( 414, 40 )[l]{$e_R$}
  \Text( 308, 32 )[c]{$g_1$}
  \Text( 376, 32 )[c]{$g_2$}
  \Text( 320, 80 )[c]{$\Wino^0$}
  \Text( 360, 80 )[c]{$\Bino$}
  \Text( 322, 28 )[c]{$\smu_L$}
  \Text( 342, 28 )[c]{$\smu_R$}
  \Text( 362, 28 )[c]{$\se_R$}
  \Text( 230, 40 )[c]{$a_{\mu e \ph}^{(o-L-\mu)}$:}
\end{picture}
\begin{picture}(440,80)
%
  \DashLine(    80, 40 )( 140, 40 ){4}
  \PhotonArc(  110, 40 )( 30, 0, 180 ){5}{8}
  \CArc(       110, 40 )( 30, 0, 180 )
  \Line(  40, 40 )( 80, 40 )
  \Line( 140, 40 )( 180, 40 )
  \Text( 160, 40 )[c]{\huge $\times$}
  \Text( 110.6, 39.8 )[c]{\huge $\bullet$}
  \Text(  36, 40 )[r]{$\mu_L$}
  \Text( 184, 40 )[l]{$\mu_R$}
  \Text(  80, 32 )[c]{$Y_\mu$}
  \Text( 140, 32 )[c]{$g_1$}
  \Text( 110, 83 )[c]{$\Higgsino_d^0/\Bino$}
  \Text( 100, 28 )[c]{$\smu_R$}
  \Text( 124, 28 )[c]{$\smu_L$}
  \Text(   0, 40 )[c]{$a_\mu^{(p-L)}$:}
%
  \DashLine(   310, 40 )( 370, 40 ){4}
  \PhotonArc(  340, 40 )( 30, 0, 180 ){5}{8}
  \CArc(       340, 40 )( 30, 0, 180 )
  \Line( 270, 40 )( 310, 40 )
  \Line( 370, 40 )( 410, 40 )
  \Text( 390, 40 )[c]{\huge $\times$}
  \Text( 329.5, 39.5 )[c]{\huge $\bullet$}
  \Text( 353.5, 40 )[c]{\large $\otimes$}
  \Text( 266, 40 )[r]{$\mu_L$}
  \Text( 414, 40 )[l]{$e_R$}
  \Text( 308, 32 )[c]{$Y_\mu$}
  \Text( 376, 32 )[c]{$g_1$}
  \Text( 340, 83 )[c]{$\Higgsino_d^0/\Bino$}
  \Text( 322, 28 )[c]{$\smu_R$}
  \Text( 342, 28 )[c]{$\smu_L$}
  \Text( 362, 28 )[c]{$\se_L$}
  \Text( 230, 40 )[c]{$a_{\mu e \ph}^{(p-L)}$:}
\end{picture}
\begin{picture}(440,80)
%
  \DashLine(    80, 40 )( 140, 40 ){4}
  \PhotonArc(  110, 40 )( 30, 0, 180 ){5}{8}
  \CArc(       110, 40 )( 30, 0, 180 )
  \Line(  40, 40 )( 80, 40 )
  \Line( 140, 40 )( 180, 40 )
  \Text(  60, 40 )[c]{\huge $\times$}
  \Text( 110.6, 39.8 )[c]{\huge $\bullet$}
  \Text(  36, 40 )[r]{$\mu_L$}
  \Text( 184, 40 )[l]{$\mu_R$}
  \Text(  80, 31 )[c]{$Y_\mu$}
  \Text( 140, 32 )[c]{$g_1$}
  \Text( 110, 83 )[c]{$\Higgsino_d^0/\Bino$}
  \Text( 100, 28 )[c]{$\smu_L$}
  \Text( 124, 28 )[c]{$\smu_R$}
  \Text(   0, 40 )[c]{$a_\mu^{(q-L)}$:}
%
  \DashLine(   310, 40 )( 370, 40 ){4}
  \PhotonArc(  340, 40 )( 30, 0, 180 ){5}{8}
  \CArc(       340, 40 )( 30, 0, 180 )
  \Line( 270, 40 )( 310, 40 )
  \Line( 370, 40 )( 410, 40 )
  \Text( 290, 40 )[c]{\huge $\times$}
  \Text( 329.5, 39.5 )[c]{\huge $\bullet$}
  \Text( 353.5, 40 )[c]{\large $\otimes$}
  \Text( 266, 40 )[r]{$\mu_L$}
  \Text( 414, 40 )[l]{$e_R$}
  \Text( 308, 31 )[c]{$Y_\mu$}
  \Text( 376, 32 )[c]{$g_1$}
  \Text( 340, 83 )[c]{$\Higgsino_d^0/\Bino$}
  \Text( 322, 28 )[c]{$\smu_L$}
  \Text( 342, 28 )[c]{$\smu_R$}
  \Text( 362, 28 )[c]{$\se_R$}
  \Text( 230, 40 )[c]{$a_{\mu e \ph}^{(q-L)}$:}
\end{picture}
\begin{picture}(440,80)
%
  \DashLine(    80, 40 )( 140, 40 ){4}
  \PhotonArc(  110, 40 )( 30, 0, 180 ){5}{8}
  \CArc(       110, 40 )( 30, 0, 180 )
  \Line(  40, 40 )( 80, 40 )
  \Line( 140, 40 )( 180, 40 )
  \Text( 160, 40 )[c]{\huge $\times$}
  \Text( 110.6, 39.8 )[c]{\huge $\bullet$}
  \Text(  36, 40 )[r]{$\mu_L$}
  \Text( 184, 40 )[l]{$\mu_R$}
  \Text(  80, 31 )[c]{$Y_\mu$}
  \Text( 140, 32 )[c]{$g_2$}
  \Text( 110, 83 )[c]{$\Wino/\Higgsino_d^0$}
  \Text( 100, 28 )[c]{$\smu_R$}
  \Text( 124, 28 )[c]{$\smu_L$}
  \Text(   0, 40 )[c]{$a_\mu^{(r-L)}$:}
%
  \DashLine(   310, 40 )( 370, 40 ){4}
  \PhotonArc(  340, 40 )( 30, 0, 180 ){5}{8}
  \CArc(       340, 40 )( 30, 0, 180 )
  \Line( 270, 40 )( 310, 40 )
  \Line( 370, 40 )( 410, 40 )
  \Text( 390, 40 )[c]{\huge $\times$}
  \Text( 329.5, 39.5 )[c]{\huge $\bullet$}
  \Text( 353.5, 40 )[c]{\large $\otimes$}
  \Text( 266, 40 )[r]{$\mu_L$}
  \Text( 414, 40 )[l]{$e_R$}
  \Text( 308, 31 )[c]{$Y_\mu$}
  \Text( 376, 32 )[c]{$g_2$}
  \Text( 340, 83 )[c]{$\Wino/\Higgsino_d^0$}
  \Text( 322, 28 )[c]{$\smu_R$}
  \Text( 342, 28 )[c]{$\smu_L$}
  \Text( 362, 28 )[c]{$\se_L$}
  \Text( 230, 40 )[c]{$a_{\mu e \ph}^{(r-L)}$:}
\end{picture}
\caption{Similar to Fig.~\ref{neut-L-fig}, the neutralino-slepton 
contributions to $a_l$ 
with a left-right slepton mass insertion
that give rise to muon $g-2$ and $\mu \ra e \ph$ in the interaction 
eigenstate basis are shown.  The slepton left-right mass insertion
is shown by a $\bullet$.}
\label{neut-L-LR-fig}
\end{figure}

Unlike the chargino-sneutrino class of diagrams, there may be
unsuppressed contribution to either $a_l$ or $a_r$.  We show
the complete set of diagrams in Figs.~\ref{neut-R-fig}--\ref{neut-L-LR-fig}.
There is a nearly one-to-one correspondence between the contribution 
to $a_l$ through left-left flavor mixing, and that for $a_r$ through 
right-right flavor mixing (or vice versa).  The exception are the 
diagrams with a Wino which couples only to left-handed fields.  

As in the chargino-sneutrino case, roughly half of the diagrams
are proportional to the electron mass or electron Yukawa coupling.  
Those diagrams that have an exact analog which simply replaces 
the electron mass with the muon mass can be safely neglected.
This includes Figs.~\ref{neut-R-fig}(e-R), \ref{neut-L-fig}(e-L), 
\ref{neut-L-fig}(g-L), \ref{neut-R-fig}(i-R), \ref{neut-L-fig}(i-L), 
\ref{neut-R-fig}(k-R), \ref{neut-L-fig}(k-L), \ref{neut-R-fig}(p-R),
\ref{neut-L-fig}(p-L), and \ref{neut-L-fig}(r-L).

Neutralino-slepton contributions include a new set of diagrams 
resulting from left-right mixing between the sleptons.  We have shown 
this set of diagrams separately in Figs.~\ref{neut-R-LR-fig} and 
\ref{neut-L-LR-fig}.  Each muon (g - 2) diagram for a given muon
chirality has \emph{two} $\mu \ra e\ph$ contributions 
that result from the two possible orderings of the left-to-right
transition and the flavor transition.  One of these orderings involves 
muon left-right mixing whereas the other involves electron left-right mixing.  
(Only the muon left-right mixing is shown in Figs.~\ref{neut-R-LR-fig} 
and \ref{neut-L-LR-fig}.)
Ordinarily the flavor-diagonal slepton mass matrix is written as
\begin{eqnarray}
{\cal M}_{\sll}^2 &=& 
    \left( \begin{array}{cc}
           m_{\sll_L}^2 & m_\ell (- \mu \tan\beta + A_\ell) \\
           m_\ell (- \mu \tan\beta + A_\ell) & m_{\sll_R}^2 
           \end{array} \right)
\end{eqnarray}
As long as the different flavor $A$-terms 
are not as hierarchically different as the lepton 
masses,
the diagrams with electron left-right mixing can be neglected.
Even though we will assume this in what follows it is clear that
weakening this restriction will not affect the bound.

The final class of diagrams to be considered are those with \emph{only} 
a Higgsino in the loop.  In (g - 2), it is straightforward to show that 
there are always larger contributions from either mixed gaugino/Higgsino 
diagrams or from pure gaugino diagrams.  We may safely neglect
the diagrams Figs.~\ref{neut-R-fig}(h-R) and \ref{neut-L-fig}(h-L) 
in favor of \ref{neut-R-fig}(j-L) and \ref{neut-L-fig}(j-R),
respectively.  This is because the Bino content of the 
lightest neutralino goes as $N_{11}\simeq \sin \theta_W M_Z/M_1$ 
(and the Wino content goes as $N_{12}\simeq \cos \theta_W M_Z/M_2$). 
The Higgsino diagram is proportional to 
$m_\mu Y_\mu^2 |N_{13}|^2$, whereas the mixed Higgsino/Bino contribution 
is proportional to $m_{\N_1} g_1 Y_\mu N_{11} N_{13}$.
So long as $\sin \theta_W M_Z/M_1 \gsim m_\mu Y_\mu/(m_{\N_1} g_1)$, 
the small Bino content of the lightest neutralino dominates the amplitude.  
This is precisely analogous to neglecting the
Higgsino diagram in favor of the mixed Higgsino/Wino diagram
for the chargino-sneutrino contribution.  

The Higgsino diagrams with left-right mixing, 
Figs.~\ref{neut-R-LR-fig}(n-R) and \ref{neut-L-LR-fig}(n-L), are 
somewhat more subtle
in the limit $\mu \ll M_1,M_2$.  In these two diagrams, the
chirality flip is on the internal line, with $\tan\beta$ enhanced
left-right 
mixing on the slepton line. However these graphs can be neglected 
relative to \ref{neut-R-LR-fig}(m-R) and \ref{neut-L-LR-fig}(m-L) 
so long as the ratio of the Bino mass to the Higgsino mass is less 
than about $(g_1/Y_{\mu})^2$.  Therefore neglecting
this diagram relative to the others is consistent in all but highly
fine-tuned regions of parameter space.

Interestingly, the remaining contributions 
to $a_r$ and $a_l$ depend exclusively on the left-left flavor changing
transition mass ${m_{12}^{LL}}^2$ and the right-right flavor changing
transition mass ${m_{12}^{RR}}^2$.  Hence, there is essentially no 
interference between the amplitudes involving a left-left flavor 
changing transition with the right-right flavor changing transition.

In general, the charged slepton mass matrix takes the form
\begin{eqnarray}
{\cal M}_{\sll}^2 &=& 
\left( \begin{array}{cccc}
       \bar{m}_{e_L}^2 & m_e (A_e - \mu \tan\beta) &
{\bar{m^2}}_{12}^{LL}
& \\
       m_e (A_e - \mu \tan\beta) & \bar{m}_{e_R}^2 & &
{\bar{m^2}_{12}^{RR}}
\\
       {\bar{m^2}_{12}^{LL}} & & \bar{m}_{\mu_R}^2 & m_\mu (A_\mu - \mu
\tan\beta) \\
       & {\bar{m^2}_{12}^{RR}} & m_\mu (A_\mu - \mu \tan\beta) &
\bar{m}_{\mu_L}^2 
       \end{array} \right) \; .
\end{eqnarray}
This $4\times4$ matrix must be diagonalized
to carry out an exact calculation.  We wish to exploit the ease
of computing in the $2\times2$ case.  
We can achieve this by first diagonalizing
the upper left and bottom right $(2\times2)$ blocks of this matrix
individually. As long as we are working in the one flavor violating
insertion limit then this problem reduces to that of several $(2\times2)$ 
matrices which can be handled independently. 

So, we decompose the $4x4$ matrix into {\footnote{Other possible 
decompositions exist, but this one suffices for the purpose of
establishing a bound. }}
\begin{eqnarray}
\renewcommand{\arraystretch}{1.3}
{\cal M}_{\sll_L}^2 \;=\; 
\left( \begin{array}{cc}
       m_{e_L}^2 & {m_{12}^{LL}}^2 \\
       {m_{12}^{LL}}^2 & m_{\mu_L}^2 
       \end{array} \right) \quad &,& \quad
{\cal M}_{\sll_R}^2 \;=\; 
\left( \begin{array}{cc}
       m_{e_R}^2 & {m_{12}^{RR}}^2 \\
       {m_{12}^{RR}}^2 & m_{\mu_R}^2 
       \end{array} \right) \; 
\end{eqnarray}  
It is important to note that the elements of these matrices are not
exactly those of the original mass matrix we started with, i.e.
$m_{e_L}$ is now not the mass of the left handed slepton in
the original mass matrix but the mass of the eigenstate which is 
`mostly' the original left handed slepton. Similarly $m_{12}^{LL}$
here is simply the mass in the corresponding position of the mass matrix
after the $(2\times2)$ blocks have been diagonalized.
The resulting matrices can be 
easily diagonalized just like the chargino-sneutrino case,
resulting in the ``mass eigenvalues'' $m_{\sll^L_{1,2}}$,
$m_{\sll^R_{1,2}}$ and mixing angles $\theta_{\sll^L}$,
$\theta_{\sll^R}$.


Using our treatment for left-right mixing, we can now write the
contribution to the (g - 2) amplitude for those diagrams with unsuppressed 
$\mu \ra e\ph$ analogs:
\begin{eqnarray}
\frac{16\pi^2}{m_\mu} a_\mu^{(d-R)} &\simeq& 
    -\frac{m_\mu}{24 m_{\N_k}^2} g_1^2 |N_{k1}|^2 
    x_{k,L} F_1^N(x_{k,L}) 
    \label{g-2-d-R-eq} \\
\frac{16\pi^2}{m_\mu} a_\mu^{(f-R)} &\simeq& 
    -\frac{m_\mu}{24 m_{\N_k}^2} g_2^2 |N_{k2}|^2 x_{k,L} F_1^N(x_{k,L}) 
    \label{g-2-f-R-eq} \\
\frac{16\pi^2}{m_\mu} a_\mu^{(g-R)} &\simeq& 
    -\frac{m_\mu}{12 m_{\N_k}^2} g_1 g_2 \mbox{Re}[N_{k1} N_{k2}^*] 
    x_{k,L} F_1^N(x_{k,L}) 
    \label{g-2-g-R-eq} \\
\frac{16\pi^2}{m_\mu} a_\mu^{(j-R)} &\simeq& 
    \frac{1}{3 \sqrt{2} m_{\N_k}} g_1 Y_\mu \mbox{Re}[N_{k1} N_{k3}] 
    x_{k,L} F_2^N(x_{k,L}) 
    \label{g-2-j-R-eq} \\
\frac{16\pi^2}{m_\mu} a_\mu^{(l-R)} &\simeq& 
    \frac{1}{3 \sqrt{2} m_{\N_k}} g_2 Y_\mu \mbox{Re}[N_{k2} N_{k3}] 
    x_{k,L} F_2^N(x_{k,L}) 
    \label{g-2-l-R-eq} \\
\frac{16\pi^2}{m_\mu} a_\mu^{(m-R)} &\simeq& 
    \frac{m_\mu (A_\mu - \mu \tan\beta)}{3 m_{\smu_R}^2 m_{\N_k}} 
    g_1^2 \mbox{Re}[N_{k1} N_{k1}] x_{k,L} F_2^N(x_{k,L}) 
    \label{g-2-m-R-eq} \\
\frac{16\pi^2}{m_\mu} a_\mu^{(o-R)} &\simeq& 
    \frac{m_\mu (A_\mu - \mu \tan\beta)}{3 m_{\smu_R}^2 m_{\N_k}} 
    g_1 g_2 \mbox{Re}[N_{k1} N_{k2}] x_{k,L} F_2^N(x_{k,L}) 
    \label{g-2-o-R-eq} \\
\frac{16\pi^2}{m_\mu} a_\mu^{(q-R)} &\simeq& 
    -\frac{\sqrt{2} m_\mu^2 (A_\mu - \mu \tan\beta)}{3 m_{\smu_R}^2 m_{\N_k}^2}
    g_1 Y_\mu \mbox{Re}[N_{k1} N_{k3}^*] x_{k,L} F_1^N(x_{k,L}) 
    \label{g-2-q-R-eq} \\
\frac{16\pi^2}{m_\mu} a_\mu^{(r-R)} &\simeq& 
    -\frac{\sqrt{2} m_\mu^2 (A_\mu - \mu \tan\beta)}{3 m_{\smu_R}^2 m_{\N_k}^2}
    g_2 Y_\mu \mbox{Re}[N_{k2} N_{k3}^*] x_{k,L} F_1^N(x_{k,L}) 
    \label{g-2-r-R-eq} 
\end{eqnarray}
and 
\begin{eqnarray}
\frac{16\pi^2}{m_\mu} a_\mu^{(d-L)} &\simeq& 
    - \frac{m_\mu}{6 m_{\N_k}^2} g_1^2 |N_{k1}|^2 
    y_{k,R} F_1^N(y_{k,R}) 
    \label{g-2-d-L-eq} \\
\frac{16\pi^2}{m_\mu} a_\mu^{(j-L)} &\simeq& 
    \frac{\sqrt{2}}{3 m_{\N_k}} g_1 Y_\mu \mbox{Re}[N_{k1} N_{k3}] 
    y_{k,R} F_2^N(y_{k,R}) 
    \label{g-2-j-L-eq} \\
\frac{16\pi^2}{m_\mu} a_\mu^{(m-L)} &\simeq& 
    \frac{\sqrt{2} m_\mu (A_\mu - \mu \tan\beta)}{3 m_{\smu_L}^2 m_{\N_k}} 
    g_1^2 \mbox{Re}[N_{k1} N_{k1}] y_{k,R} F_2^N(y_{k,R}) 
    \label{g-2-m-L-eq} \\
\frac{16\pi^2}{m_\mu} a_\mu^{(o-L)} &\simeq& 
    \frac{\sqrt{2} m_\mu (A_\mu - \mu \tan\beta)}{3 m_{\smu_L}^2 m_{\N_k}} 
    g_1 g_2 \mbox{Re}[N_{k1} N_{k2}] y_{k,R} F_2^N(y_{k,R}) \; .
    \label{g-2-o-L-eq} \\
\frac{16\pi^2}{m_\mu} a_\mu^{(q-L)} &\simeq& 
  -\frac{2 \sqrt{2} m_\mu^2 (A_\mu - \mu \tan\beta)}{3 m_{\smu_L}^2 m_{\N_k}^2}
    g_1 Y_\mu \mbox{Re}[N_{k1} N_{k3}^*] y_{k,R} F_1^N(y_{k,R}) 
    \label{g-2-q-L-eq}
\end{eqnarray}
where the sum over $k=1,2,3,4$ for the four neutralinos is 
implicitly understood.  Here $g_1$ is the U(1)$_Y$ coupling,
$x_{k,L} \equiv m_{\N_k}^2/m_{\smu_L}^2$, 
$y_{k,R} \equiv m_{\N_k}^2/m_{\smu_R}^2$, 
and $N_{ij}$ is the neutralino mixing matrix in the
$(\Bino,\Wino^0,\Higgsino_d^0,\Higgsino_u^0)$ basis.
The one-loop kinematical functions $F(x)$ are defined identically 
to Ref.~\cite{MW} and are given in the Appendix.

The contribution to $\mu \ra e\ph$ can be obtained from
the above amplitudes by replacing 
\begin{eqnarray}
x_{k,L} F_i^N(x_{k,L}) &\longrightarrow& \frac{1}{2} \sin 2\theta_{\sll^L}
    \left( x_{k,1} F_i^N(x_{k,1}) - x_{k,2} F_i^N(x_{k,2}) \right) \\
y_{k,R} F_i^N(y_{k,R}) &\longrightarrow& \frac{1}{2} \sin 2\theta_{\sll^R}
    \left( y_{k,1} F_i^N(y_{k,1}) - y_{k,2} F_i^N(y_{k,2}) \right) \; ,
\end{eqnarray}
where 
$x_{k,\{1,2\}} \equiv m_{\N_k}^2/m_{\sll^L_{\{1,2\}}}^2$ and
$y_{k,\{1,2\}} \equiv m_{\N_k}^2/m_{\sll^R_{\{1,2\}}}^2$.
This is completely analogous to the chargino-sneutrino result.
In fact, the ratio of any pair of diagrams for a given neutralino
(fixed $k$) is the same:
\begin{eqnarray}
\frac{a_{\mu e \ph}^{(R)}}{a_\mu^{(R)}} &=& 
    \frac{1}{2} \sin 2\theta_{\sll^L} \frac{x_{k,1} F_i^N(x_{k,1}) 
    - x_{k,2} F_i^N(x_{k,2})}{x_{k,L} F_i^N(x_{k,L})} \\
\frac{a_{\mu e \ph}^{(L)}}{a_\mu^{(L)}} &=& 
    \frac{1}{2} \sin 2\theta_{\sll^R} \frac{y_{k,1} F_i^N(y_{k,1}) 
    - y_{k,2} F_i^N(y_{k,2})}{y_{k,R} F_i^N(y_{k,R})}
\end{eqnarray}

Following the same arguments used the chargino-sneutrino case,
we obtain
\begin{eqnarray}
\frac{a_{\mu e \ph}^{(R)}}{a_\mu^{(R)}} &\simeq& 
\frac{{m_{12}^{LL}}^2}{\mbox{Max}[m_{\smu_L}^2,m_{\se_L}^2,m_{\N_k}^2]} \\
\frac{a_{\mu e \ph}^{(L)}}{a_\mu^{(L)}} &\simeq& 
\frac{{m_{12}^{RR}}^2}{\mbox{Max}[m_{\smu_R}^2,m_{\se_R}^2,m_{\N_k}^2]} \; .
\end{eqnarray}
Just as in the chargino-sneutrino case, the ratio can be calculated 
exactly in various limits shown in Table~\ref{neut-table}.
\begin{table}[t]
\renewcommand{\arraystretch}{1.5}\small\normalsize
\setlength{\tabcolsep}{1cm}
\begin{center}
\begin{tabular}{ccc} \hline\hline
case & $a_{\mu e \ph}/a_\mu$ (set 1) & $a_{\mu e \ph}/a_\mu$ (set 2)
   \\ \hline
$x_{k,1},x_{k,2} \gg 1$           & $-\textfrac{11}{2} m_{12}^2/m_{\N_k}^2$ &
    $-3 m_{12}^2/m_{\N_k}^2$ \\
$x_{k,1},x_{k,2} \ll 1$           & $m_{12}^2/m_{\se^L}^2$ & 
    $m_{12}^2/m_{\se^L}^2$ \\
$x_{k,1} \ll 1 \ll x_{k,2}$       & $m_{12}^2/m_{\se^L}^2$ & 
    $m_{12}^2/m_{\se^L}^2$ \\
$x_{k,2} \ll 1 \ll x_{k,1}$       & $2 m_{12}^2/m_{\N_k}^2$ & 
    $m_{12}^2/m_{\N_k}^2$ \\
$x_{k,1} \sim x_{k,2} \sim 1$     & $\textfrac{3}{5} m_{12}^2/m_{\se^L}^2$ & 
    $\textfrac{1}{2} m_{12}^2/m_{\se^L}^2$ \\
$x_{k,1} \sim 1$, $x_{k,2} \gg 1$ & $\textfrac{3}{4} m_{12}^2/m_{\se^L}^2$ & 
    $\textfrac{2}{3} m_{12}^2/m_{\se^L}^2$ \\
$x_{k,1} \sim 1$, $x_{k,2} \ll 1$ & $\textfrac{1}{2} m_{12}^2/m_{\N_k}^2$ & 
    $\textfrac{1}{3} m_{12}^2/m_{\N_k}^2$ \\
$x_{k,2} \sim 1$, $x_{k,1} \gg 1$ & $3 m_{12}^2/m_{\N_k}^2$ & 
    $2 m_{12}^2/m_{\N_k}^2$ \\
$x_{k,2} \sim 1$, $x_{k,1} \ll 1$ & $m_{12}^2/m_{\se^L}^2$ & 
    $m_{12}^2/m_{\se^L}^2$ \\
\hline\hline
\end{tabular}
\end{center}
\caption{The ratio of the amplitude for $\mu \ra e\ph$ over (g - 2),
for a given neutralino (fixed $k$), for two classes of diagrams.
Set 1:  Fig.~\ref{neut-R-fig}(d-R),(f-R),(g-R), and Fig 4(q-R),(r-R)
and Set 2:
Fig.~\ref{neut-R-fig}(j-R),(l-R) and \ref{neut-R-LR-fig}(m-R),(o-R).
The same results hold for $x \leftrightarrow y$
$m_{\sll^{L}}^2 \leftrightarrow m_{\sll^{R}}^2$, 
and ${m_{12}^{LL}}^2 \leftrightarrow {m_{12}^{RR}}^2$,
where Set 1 is just Fig.~\ref{neut-L-fig}(d-L), Fig. 6(q-L) and
Set 2 includes Fig.~\ref{neut-L-fig}(j-L), 
Fig.~\ref{neut-L-LR-fig}(m-L) and (o-L).}
\label{neut-table}
\end{table}
The results in the table closely match the expectation
above.  From this, we can predict the size of the amplitude
for $\mu \ra e\ph$ assuming that muon (g - 2) is dominated by one 
diagram (with one neutralino). Either
\begin{equation}
a_{\mu e \ph}^r \gsim \frac{1}{3}a_\mu \delta_{12}^{LL} \\
\end{equation}
{\bf{or}}
\begin{equation}
a_{\mu e \ph}^l \gsim \frac{1}{3} a_\mu \delta_{12}^{RR} \; ,
\end{equation}

Following the same analysis in the chargino-sneutrino sector, we
can use the current experimental bound on the branching ratio
for $\mu \ra e\ph$ to obtain a bound on the flavor mixing 
(mass)$^2$:
\begin{eqnarray}
\delta_{12}^{LL} < 1.8 \times 10^{-4} 
    \left( \frac{\mbox{BR}(\mu \ra e\ph)}{1.2 \times 10^{-11}} \right)^{1/2} 
    \left( \frac{4.3 \times 10^{-9}}{a_\mu} \right)
\end{eqnarray}
or
\begin{eqnarray}
\delta_{12}^{RR} < 1.8 \times 10^{-4} 
    \left( \frac{\mbox{BR}(\mu \ra e\ph)}{1.2 \times 10^{-11}} \right)^{1/2} 
    \left( \frac{4.3 \times 10^{-9}}{a_\mu} \right)
\end{eqnarray}


Combining the chargino-sneutrino results with the neutralino-slepton
results, we obtain roughly the same result given above.  
Notice that, without making assumptions about the soft mass
hierarchy, the best we can do is to place a bound on $\delta_{12}^{LL}$ 
\emph{or} $\delta_{12}^{RR}$, but not both simultaneously.

\section{Other lepton flavor violating decays}
\label{tau-sec}

We now consider constraints on the flavor violating transition 
$\tau \ra \mu \gamma$. Once again we can establish a correspondence 
between the various diagrams. Diagrams which have a chirality flip on
the external $\tau$ line or have a tau Yukawa vertex have an 
$m_{\tau}/m_{\mu}$ amplitude enhancement which is not there in the 
$\mu \ra e \gamma$ decay. The (g - 2) graphs can be written in
pairs, with the same particles in the loop but
the ingoing and outgoing muons having different chiralities. At least
one of the graphs in each pair is always enhanced with the exception of 
[${a_{\mu}}^{m-R}$ and ${a_{\mu}}^{m-L}$], [${a_{\mu}}^{o-R}$ and
${a_{\mu}}^{o-L}$]. These graphs are not enhanced unless we assume 
that $A_{l}$ is generation independent and hence the off diagonal
elements in the slepton mass matrix are proportional to the Yukawa
couplings. From now on we will assume that this is the case and
proceed. Writing the relevant part of the interaction Lagrangian as  
\begin{equation}
{\cal M} = \frac{i e}{2 m_\mu}
             \overline{u}_\mu(p_2) \sigma^{\mu\nu} q_{\nu}
             \left( a_l P_L + a_r P_R \right) u_\tau(p_1) A_\mu
             + {\rm{h.c.}} 
\end{equation}
In analogy with the $\mu \ra e \gamma$ 
the smallest value of $a_l$ or $a_r$ corresponding
to this process is approximately given by 
\begin{equation}
a =
\frac{1}{4}a_{\mu}\frac{m_{\tau}}{m_{\mu}}\frac{{m_{23}}^2}{{\bar{m}}^2}
\end{equation}
The width is given by
\begin{eqnarray}
\Gamma(\mu \ra e\ph) &=& \frac{{m_\tau}^3 e^2}{64\pi {m_{\mu}}^2}
                         \left( |a_l|^2 + |a_r|^2 \right) \; .     
\end{eqnarray}

Using the experimental bound on the branching ratio ($ \le 1.1 \times
10^{-6}$) \cite{Ahmed} and from the the lifetime of the $\tau$ ( $2.9
\times
10^{-13}$s) we obtain 
\begin{equation}
\frac{{m_{23}}^2}{{\bar{m}}^2} \le 1.4 \times 10^{-1}
\end{equation}

\section{Conclusions}
\label{conclusions-sec}

We have found a precise correspondence between the supersymmetric
diagrams that contribute to the muon anomalous magnetic moment 
and those which contribute to the flavor violating processes $\mu \ra e\ph$
and $\tau \ra \mu\ph$.  Using current experimental limits on the
branching ratios of these decay modes, combined with the assumption of a
supersymmetric contribution to the muon anomalous magnetic moment, 
we have found strong bounds on the size of the $e \leftrightarrow \mu$ 
lepton flavor violating soft mass, essentially independent of assumptions
other supersymmetric parameters.  Assuming the current deviation measured 
at BNL is from supersymmetry, and using the current experimental
limits on radiative leptonic branching ratios, 
we find ${m^2}_{e \mu}/ {{\bar{m}}^2} \lsim 2 \times 10^{-4}$ and
${m^2}_{\tau \mu}/ {{\bar{m}}^2} \lsim  1 \times 10^{-1}$, where
${\bar{m}}^2$ is the mass of the heaviest particle in any loop that
contributes at this level to the anomalous magnetic moment of the muon.
Improvement in 
the experimental measurement of (g - 2) can be easily incorporated
into our results since our bound is inversely proportional to the 
central value discrepancy between the standard model and experiment.

The absence of lepton flavor violation places a significant constraint
on the non-flavor-blind mediation of supersymmetry breaking that often
occurs at a suppressed level in many models.  Finding lepton flavor
violation, however, could lead to fascinating ways of accessing
aspects of supersymmetry breaking models that are not easily
obtained through other means, such as estimating the size of
the extra dimension in anomaly mediation or gaugino mediation models.

\emph{Note added:}  As this paper was being completed, another
paper recently appeared \cite{GraesserThomas} that also discussed
the connection between lepton flavor violation and muon (g - 2), 
with similar conclusions.

\section*{Acknowledgments}
\indent

We thank A. Nelson for useful discussions.
We also thank the the theoretical physics group at LBL for a stimulating
atmosphere where this work was initiated.
This work was supported in part by the U.S. Department of Energy 
under grant numbers DE-FG03-96-ER40956 and DE-FG02-95-ER40896.

\begin{appendix}

\section*{Appendix:  One-loop kinematical functions and their limits.}

The kinematical functions $F(x)$ that arise in muon (g - 2) and
the lepton flavor violating processes $\ell_i \ra \ell_j\ph$
are given by \cite{MW}
\begin{eqnarray}
F_1^C(x) &=& \frac{2}{(1 - x)^4} 
             \left[ 2 + 3 x - 6 x^2 + x^3 + 6 x \ln x \right] \\
F_2^C(x) &=& -\frac{3}{2 (1 - x)^3} 
             \left[ 3 - 4 x + x^2 + 2 \ln x \right] \\
F_1^N(x) &=& \frac{2}{(1 - x)^4} 
             \left[ 1 - 6 x + 3 x^2 + 2 x^3 - 6 x^2 \ln x \right] \\
F_2^N(x) &=& \frac{3}{(1 - x)^3} 
             \left[ 1 - x^2 + 2 x \ln x \right] \; .
\end{eqnarray}
There are three interesting regions of these functions:
small $x$, $x \sim 1$, and large $x$.  In the small $x$ limit, 
$x F(x)$ can be written as
\begin{eqnarray}
x F_1^C(x) &=& 4 x + 22 x^2 + \ldots \\
x F_2^C(x) &=& 3 x \ln 1/x - \frac{9}{2} x - \frac{15}{2} x^2 + \ldots \\
x F_1^N(x) &=& 2 x - 4 x^2 + \ldots \\
x F_2^N(x) &=& 3 x + 9 x^2 + \ldots \quad .
\end{eqnarray}
For $x \sim 1$, $x F(x)$ can be written as
\begin{eqnarray}
x F_1^C(x) &=& 1 + \frac{2}{5} (x - 1) - \frac{1}{5} (x - 1)^2 + \ldots \\
x F_2^C(x) &=& 1 + \frac{1}{4} (x - 1) - \frac{3}{20} (x - 1)^2 + \ldots \\
x F_1^N(x) &=& 1 + \frac{3}{5} (x - 1) - \frac{1}{5} (x - 1)^2 + \ldots \\
x F_2^N(x) &=& 1 + \frac{1}{2} (x - 1) - \frac{1}{5} (x - 1)^2 + \ldots \quad .
\end{eqnarray}
Finally, in the large $x$ limit, $x F(x)$ can be written as
\begin{eqnarray}
x F_1^C(x) &=& 2 - \frac{4}{x} - \frac{22}{x^2} + \ldots \\
x F_2^C(x) &=& \frac{3}{2} - \frac{3}{2 x} - \frac{9}{2 x^2} + \ldots \\
x F_1^N(x) &=& 4 + \frac{22}{x} + \frac{52}{x^2} + \ldots \\
x F_2^N(x) &=& 3 + \frac{9}{x} + \frac{15}{x^2} + \ldots \quad .
\end{eqnarray}

\end{appendix}


\newpage

\newpage

\end{document}